\documentclass[acmsmall,screen]{acmart}
\usepackage{algorithm2e}
\RestyleAlgo{ruled}
\usepackage{multirow}
\usepackage{enumitem}

\newcommand{\name}{OBsmith}

\usepackage{listings, xcolor}
\definecolor{javapurple}{rgb}{0.5,0,0.35} 
\definecolor{linenumbergray}{rgb}{0.5,0.5,0.5}
\lstdefinestyle{Java-github}{
        basicstyle=\ttfamily\scriptsize,
        language=Java,
        commentstyle=\color{linenumbergray},
        stringstyle=\color{javapurple},
        keywordstyle=\color{red},
        morekeywords={@Test},
        morecomment=[s][\color{linenumbergray}]{/**}{*/},
        numbers=left,
        numberstyle=\tiny\color{linenumbergray},
        numbersep=2.5pt,
        xleftmargin=1em,
        moredelim=**[is][\color{javapurple}]{@h@}{@h@},
        morecomment=[f][{\btHL[fill=gitdel]}]-,
        morecomment=[f][{\btHL[fill=gitadd]}]+,
        breaklines = true,
}

\lstdefinestyle{obf}{
        basicstyle=\ttfamily\footnotesize,
        language=Html,
        commentstyle=\color{linenumbergray},
        stringstyle=\color{javapurple},
        keywordstyle=\color{red},
        morekeywords={@Test},
        morecomment=[s][\color{linenumbergray}]{/**}{*/},
        numberstyle=\tiny\color{linenumbergray},
        numbersep=2.5pt,
        xleftmargin=1em,
        moredelim=**[is][\color{javapurple}]{@h@}{@h@},
        morecomment=[f][{\btHL[fill=gitdel]}]-,
        morecomment=[f][{\btHL[fill=gitadd]}]+,
        breaklines = true,
        frame=single
}

\lstdefinestyle{prompt}{
        basicstyle=\ttfamily\scriptsize,
        language=Html,
        commentstyle=\color{linenumbergray},
        stringstyle=\color{javapurple},
        keywordstyle=\color{red},
        morekeywords={@Test},
        morecomment=[s][\color{linenumbergray}]{/**}{*/},
        numberstyle=\tiny\color{linenumbergray},
        numbersep=2.5pt,
        xleftmargin=1em,
        moredelim=**[is][\color{javapurple}]{@h@}{@h@},
        morecomment=[f][{\btHL[fill=gitdel]}]-,
        morecomment=[f][{\btHL[fill=gitadd]}]+,
        breaklines = true,
        frame=single
}

\AtBeginDocument{%
  }

\setcopyright{rightsretained}
\acmDOI{10.1145/3798204}
\acmYear{2026}
\acmJournal{PACMPL}
\acmVolume{10}
\acmNumber{OOPSLA1}
\acmArticle{96}
\acmMonth{4}
\received{2025-10-10}
\received[accepted]{2026-02-17}

\begin{document}

\title{OBsmith: LLM-Powered JavaScript Obfuscator Testing}

\author{Shan Jiang}
\orcid{0009-0002-0466-1028}
\affiliation{%
  \institution{University of Texas at Austin}
  \city{Austin}
  \country{USA}
}
\email{shanjiang@utexas.edu}

\author{Chenguang Zhu}
\orcid{0000-0002-7343-8279}
\affiliation{%
  \institution{University of Texas at Austin}
  \city{Austin}
  \country{USA}
}
\email{cgzhu@utexas.edu}

\author{Sarfraz Khurshid}
\orcid{0009-0009-7424-7819}
\affiliation{%
  \institution{University of Texas at Austin}
  \city{Austin}
  \country{USA}
}
\email{khurshid@ece.utexas.edu}

\renewcommand{\shortauthors}{Shan et al.}

\newcommand{\Comment}[1]{}

\begin{abstract}
JavaScript obfuscators are widely deployed to protect intellectual property and resist reverse engineering, yet their correctness has been largely overlooked compared to performance and resilience. Existing evaluations typically measure resistance to deobfuscation, leaving the critical question of whether obfuscators preserve program semantics unanswered. Incorrect transformations can silently alter functionality, compromise reliability, and erode security—undermining the very purpose of obfuscation. To address this gap, we present OBsmith, a novel framework to systematically test JavaScript obfuscators using large language models (LLMs). OBsmith leverages LLMs to generate program sketches—abstract templates capturing diverse language constructs, idioms, and corner cases—which are instantiated into executable programs and subjected to obfuscation under different configurations. Besides LLM-powered sketching, OBsmith also employs a second source: automatic extraction of skeletons from real programs. This extraction path enables more focused testing of project-specific features and lets developers inject domain knowledge into the resulting test cases. OBsmith uses two techniques to derive test oracles: \textit{(i)} reference-oriented equivalence testing, which takes the \emph{original program as reference oracle (ground truth)} and checks whether the obfuscated version preserves equivalent functionality, and \textit{(ii)} metamorphic testing, which applies semantics-preserving transformations to the original program and checks if obfuscation violates expected behavior.

We evaluate OBsmith on two widely used obfuscators, Obfuscator.IO and JS-Confuser, generating 600 sketches using six popular LLMs. OBsmith fills these sketches and generates over 3,000 candidate programs and obfuscates them across seven obfuscation configurations. OBsmith uncovers 11 previously unknown correctness bugs. Under an equal program budget, five general purpose state-of-the-art JavaScript fuzzers (FuzzJIT, Jsfunfuzz, Superion, DIE, Fuzzilli) failed to detect these issues, highlighting OBsmith’s complementary focus on obfuscation-induced misbehavior. An ablation shows that all components except our generic MRs contribute to at least one bug class; the negative MR result suggests the need for obfuscator-specific metamorphic relations. 
Our results also seed a discussion on how to balance obfuscation presets and performance cost. We envision OBsmith as an important step towards automated testing and quality assurance of obfuscators and other semantic-preserving toolchains.
\end{abstract}



\maketitle

\newcommand{\llm}{6}
\newcommand{\sketch}{60}
\newcommand{\obf}{2}
\newcommand{\bug}{6}

\section{Introduction}
Software obfuscation is an important technique that makes code difficult to read, analyze, or reverse engineer. This objective is typically achieved through code transformations such as changing code structures, renaming variables to non-descriptive identifiers, introducing misleading or redundant code, restructuring control flows, and encrypting specific data and strings \cite{6185286,zhang2021android,canfora2015obfuscation}. While obfuscation can serve legitimate purposes, including the protection of proprietary code or sensitive information such as IP addresses \cite{doyle2018privacy,lynn2004positive}, it is also widely adopted by malicious actors to conceal the intent of malicious scripts, particularly within web and Android applications \cite{brezinski2023metamorphic,281380,10.1016/j.cose.2015.02.007,meng2023smartphone}. JavaScript, the predominant language for client-side web applications, is particularly susceptible to obfuscation due to its open and accessible nature \cite{fraunholz2018defending,brewer2010link,10.5555/2028067.2028070,meng2022batmapper}. Any JavaScript code executed within a browser can be easily viewed and analyzed, rendering it an attractive target for both benign and malicious obfuscation. JavaScript is a versatile language, functioning across both client-side and server-side environments, each characterized by distinct execution contexts, capabilities, and security requirements.

By design, an effective obfuscator is expected to complicate the original source code without altering its intended functionality. The primary goal is to make the code substantially more difficult for reverse engineers and automated analysis tools to understand, thus safeguarding proprietary or sensitive information. Ensuring functional equivalence between obfuscated and original code is critical, as obfuscation may introduce unintended behaviors, potentially undermining software reliability or security. There is still a lack of research on evaluating obfuscators on pure JavaScript code. Previous research on JavaScript obfuscation largely focuses on evaluating the effectiveness of obfuscation techniques or detecting if code is obfuscated. However, comprehensive testing frameworks that validate both functional correctness and runtime performance impacts of JavaScript obfuscators remain underexplored.  Recent research by Skolka et al. \cite{skolka2019anything} offers a comprehensive evaluation of JavaScript minifiers and obfuscators, with particular emphasis on their robustness when applied to client-side scripts. The authors primarily relied on unit tests to assess the effectiveness of various obfuscators. Their findings reveal that these tools often exhibit insufficient resilience. A core challenge of client-side script is that client-side JavaScript is inherently dependent on complex factors such as user interactions, asynchronous events, and integration with real-world browser APIs (e.g., DOM, cookies, localStorage). These dependencies are difficult to replicate or comprehensively test using automated methods. As a result, automated source code transformations—such as minification and obfuscation—may inadvertently introduce bugs in client-side. Moreover, prior studies have omitted the evaluation of some of the most widely adopted obfuscators, such as Obfuscator.IO \cite{obio} and JS-Confuser \cite{jsconfuser}, as indicated by a recent Google survey \cite{jiang2025cascade}. This omission constitutes another important research gap, particularly given the pervasive use of these tools in the software industry. To address this limitation, the present study aims to instrument and evaluate pure JavaScript code that is executable in both Node.js and browser environments. Addressing this research gap is crucial to ensuring that JavaScript obfuscators can reliably achieve their intended security and functionality objectives without inadvertently compromising software behavior.

General purpose compiler testing approaches lack a systematically-enhanced test oracle to capture the JavaScript obfuscator bugs. Off-the-shelf compiler fuzzers assume closed-world, deterministic pipelines whose goal is transparent semantics preservation; they rely on crash oracles or n-version agreement (e.g., GCC vs.\ Clang). JavaScript obfuscators instead \emph{preserve while disguising} semantics, inject seed-dependent randomness and self-defense checks, and interact with host features that general fuzzers neither generate nor control. Bugs often surface in JS-specific corners—eval function, dynamic code loading, scoping/hoisting, prototype mutations, getters/setters, proxies—combined with obfuscation patterns like control-flow flattening or lazy string decoding. These properties defeat differential comparisons across tools and make crashes a poor proxy for silent miscompilations. Therefore, we introduce reference-oriented equivalence testing, which takes the \emph{original program as reference oracle (ground truth)} and checks whether the obfuscated version preserves equivalent functionality. Checked correctness includes return values, exceptions, scheduling-sensitive effects, and observable host-API interactions. \name~generates JS-centric, event-aware sketches, controls nondeterminism via seed management, executes in sandboxed Node and browser-like environments, and is configuration-aware to sweep obfuscator options. Lightweight instrumentation records variable states and control-flow to localize divergences. This semantics-, host-, and transformation-aware design exposes obfuscation-specific bugs that general compiler fuzzers systematically miss. Obfuscators differ from compilers because they preserve but disguise semantics. They often use tricks (control-flow flattening, dead code) that do not appear in compiler pipelines. Hence, domain-specific sketching/testing is required.

We propose \name, a novel testing framework specifically designed for evaluating JavaScript obfuscators. \name~leverages sketching techniques to systematically generate diverse test programs and employs reference-oriented equivalence testing to identify discrepancies introduced by obfuscation processes. Sketching and differential testing methodologies have previously demonstrated effectiveness in compiler testing, enabling the detection of subtle semantic bugs and inconsistencies. Extensive literature on compiler testing underscores the value and efficacy of integrating domain-specific knowledge into test-case generation. Inspired by these insights, we adopt sketching to incorporate JavaScript-specific domain knowledge into our framework, thereby enhancing the capability of \name~to detect functional deviations and performance degradations caused by obfuscators.

\name~uses LLM to generate sketches and uses Babel \cite{babel} to implement the  reference-oriented equivalence testing mechanism. Trained on extensive textual corpora, LLMs inherently encapsulate rich domain-specific knowledge, making them particularly suitable for generating sketches in automated test-case generation. Leveraging their ability to understand syntactic structures and semantic constraints, LLMs can systematically produce representative code sketches that reflect realistic programming patterns and corner cases. Integrating LLM-based sketch generation into \name~frameworks enhances test diversity and coverage with low resource, thus enabling the effective discovery of subtle functional inconsistencies and edge-case defects in software, such as those potentially introduced by JavaScript obfuscators.

A critical component of \name~is its integrated program generator, which is engineered to rigorously explore different execution paths and expose the semantic equivalence between the original source program and its obfuscated counterpart. This step is indispensable, as the intricate nature of obfuscation transformations carries an inherent risk of introducing functional regressions or altering program behavior in subtle ways. To mitigate this risk, OBsmith contains a sketch filling algorithm to generate a comprehensive set of inputs to test different execution paths. OBsmith uses a program enhancer to record global and local variables and the program's control flow. For each input, \name~asserts that the original and obfuscated program versions produce identical outputs and exhibit equivalent observable side-effects. This process provides a strong, empirical guarantee that the core logic and functionality of the application remain unaltered post-obfuscation, thereby enhancing the trustworthiness and practical deployability of our approach.

Unlike the traditional differential testing setup used in compiler testing -- where multiple variants of a system are run on the same input and their outputs compared for inconsistencies -- \name~ relies on a reference oracle to serve as ground truth. In particular, we treat the original, unobfuscated program as the reference oracle for correct behavior.
This direct comparison against a known-correct reference, termed  reference-oriented equivalence testing, allows for a precise and conclusive assessment of the correctness of the obfuscation transformations, ensuring that they have not introduced functional changes.

We evaluate OBsmith on two widely used obfuscators, Obfuscator.IO \cite{obio} and JS-Confuser \cite{jsconfuser}, generating 600 sketches using six popular LLMs. OBsmith fills these sketches and generates over 3,000 candidate programs and obfuscates them across seven obfuscation configurations. OBsmith uncovers 11 previously unknown correctness bugs. Under an equal program budget, five general purpose state-of-the-art JavaScript fuzzers (FuzzJIT, Jsfunfuzz, Superion, DIE, Fuzzilli) failed to rediscover these issues, highlighting OBsmith’s complementary focus on obfuscation-induced misbehavior. We also discuss obfuscation affect on file size, run time, and memory usage in discussion (\S\ref{sec:discussion}).

To summarize, this paper makes the following contributions:
\begin{itemize}

\item \textbf{Novelty.} This paper introduces OBsmith, the first LLM-powered framework for systematically testing JavaScript obfuscators. OBsmith introduces an LLM-driven, sketch-based generator (mixing LLM-created and automatically extracted sketches) plus a PL-aware oracle that combines  reference-oriented equivalence testing  of outputs/exceptions/termination with lightweight instrumentation and exploratory metamorphic tests. 

\item \textbf{Implementation.} We implement OBsmith with 2 sketch sources: (i) leverages multi-agent LLM system to automatically generate diverse sketches and (ii) uses existing JavaScript programs to automatically extract sketches. To solve the oracle problem, OBsmith applies reference-oriented equivalence testing and metamorphic testing to evaluate the correctness of obfuscation.

\item \textbf{Real-world Bugs.} We conduct a comprehensive study of two widely used obfuscators across seven configurations and uncover 11 correctness bugs, including silent miscompilations where the obfuscated code changes behavior. All bugs were confirmed by reproducing them in both online and repository versions. These findings highlight the unreliability of obfuscators and provide actionable insights for both developers (to fix issues) and practitioners (to make informed tool choices).

\end{itemize}

\section{Sketch Example}
\label{sec:sketch_example}
\begin{figure*}[htbp]
	\centering
	\begin{minipage}[t]{0.33\linewidth}
		\begin{lstlisting}[style = Java-github]
let x = NumberLiteral;
let y = NumberLiteral;
let text = "x*y";
console.log(NumberReference)
console.log(text)
let result = eval(text);
console.log(result + 
NumberReference);
console.log(text + 
NumberReference);
            \end{lstlisting}
	\end{minipage}
    \begin{minipage}[t]{0.3\linewidth}
		\begin{lstlisting}[style = Java-github]
let x = 10;
let y = 20;
let text = "x*y";
console.log(x)
console.log(text)
let result = eval(text);
console.log(result + x);
console.log(text + y);
            \end{lstlisting}
	\end{minipage}
    \hspace{.05in}
    \begin{minipage}[t]{0.33\linewidth}
		\begin{lstlisting}[style = Java-github]
10
x * y
undefined: 1
x * y
^    

ReferenceError: 
x is not defined
...
            \end{lstlisting}
	\end{minipage}
    \caption{A simplified sketch with holes (left), the corresponding concrete program that is input to obfuscators (middle), and the output of obfuscated program created by JS-Confuser which shows its faulty behavior (right).}
		\label{fig:sketch}
\end{figure*}

In this paper, a sketch refers to a “program with holes,” where key components - such as expressions, variables, or literals - are left as placeholders to be filled in later.  Sketches, also known as “skeletal programs” or “templates,” have been shown in previous work to be an effective mechanism for introducing domain knowledge into compiler testing \cite{jattack, Skeletal2017, WangETALABZ2018ASketch}. By abstracting certain program elements, sketches allow for a flexible yet systematic exploration of diverse code scenarios that are effective in exposing compiler bugs. The left part of Fig. 1 is an example of sketch with holes (e.g. NumberLiteral, NumberReference) and the middle is the corresponding filled code.

Building upon the sketch definition, OBsmith introduces LLMs to the sketch generation pipeline, uses the power of LLMs to automatically generate program sketches. LLMs, trained on large code corpora, inherently encode a broad range of programming concepts and domain-specific patterns. This capability allows them to produce high-quality sketches that capture the syntax and structure of real-world programs. By leveraging LLMs for sketch generation, OBsmith automates a previously manual and expertise-driven task, making it possible to efficiently create diverse and domain-informed templates. Afterward, OBsmith uses a program generator to solve the LLM-generated sketches with concrete values or variables, producing executable JavaScript programs that remain faithful to the intended domain constraints. 

OBsmith combines LLM-based sketch generation with a program generator, it enables precise and effective testing while reducing manual effort and minimizing the risk of introducing unintended errors during test case generation. By elevating sketches to the core of its workflow and integrating LLMs as a key component, OBsmith sets a new direction for leveraging LLM's domain knowledge in automated testing.

JavaScript obfuscators take two inputs: the program under obfuscation and the obfuscation configuration. Once we obfuscate the concrete program with different obfuscators and different configurations, we got the multiple obfuscated programs. Then we conduct a  reference-oriented equivalence testing on these programs to compare if they have the same functionality. Unlike traditional differential testing in compiler testing, we have ground truth (unobfuscated program). Here the concrete program serve as ground truth and we compare the running result to find bugs. Furthermore, OBsmith enhances programs by recording the program's workflow and data flow, which effectively reflects the program functionality and enables us to easily implement reference-oriented equivalence testing. An example of found bug is shown in the right part of Fig. \ref{fig:sketch}. Here the variable x is reported as undefined during execution (ReferenceError: x is not defined). This error highlights a critical functionality bug introduced by JS-Confuser, demonstrating how \name~effectively detects discrepancies between original program and obfuscated program.

\section{OBsmith Approach}
\begin{figure}[htbp]
    \centering
    \includegraphics[width=.3\linewidth]{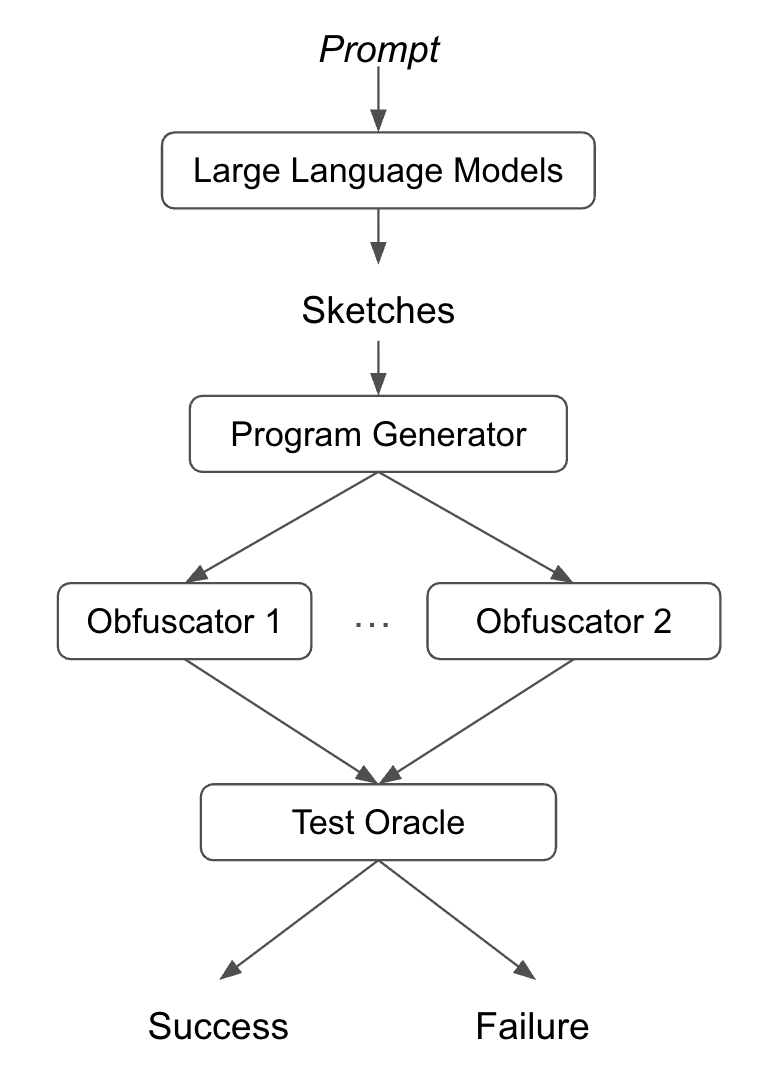}
    \caption{OBsmith overall workflow}
    \label{fig:overview}
\end{figure}
In this section, we present \name, an automated framework that leverages LLMs to generate sketches and use sketches to test JavaScript obfuscators. \name~constructs a multi-step pipeline to test JavaScript obfuscators, its overall workflow is shown in Fig. \ref{fig:overview}. 

The first step is \textbf{LLM-powered sketch generation} (\S\ref{sec:llm-sketching} and \S\ref{sec:feedback_loop}). \name~contains an initial prompt to define the sketch syntax and instruct LLMs to produce diverse and representative JavaScript sketches. \name~also contains a feedback loop to help LLM generate high quality sketches. 
Moreover, \name~has a extraction framework (\S\ref{sec:sketch_extraction}) to automatically extract sketches from existing JavaScript programs, which makes it more powerful. 

Then LLM-generated and extracted sketches are used in \textbf{program generation} (\S\ref{sec:program-generation}) to get concrete JavaScript programs. In program generation stage, \name~\textit{(i)} fills the sketch by filling all placeholders and generate concrete expressions in the sketch; and \textit{(ii)} enhances the filled program to log the program's control flow and data flow. The output of program generation stage is a set of enhanced candidate programs ready for obfuscation. 

The last step is use candidate programs and call obfuscators to get multiple obfuscated variants. These variants with corresponding ground truth are inputs to the follow-up testing stage. OBsmith uses a reference oracle for reference-oriented equivalence testing (\S\ref{sec:diff}), complemented by metamorphic testing (\S\ref{sec:metamorphic_testing}) for robustness. \name~uses different JavaScript obfuscators with different obfuscation configurations to obfuscate candidate programs. The original candidate program and all obfuscated versions are executed within a standard JavaScript engine (Node.js). Unlike traditional differential testing in compiler testing which lacks ground truth, \name~uses the original program as ground truth. \name~collects the execution outputs of all obfuscated programs and compares these outputs with the ground truth. If \name~finds any output inconsistencies, it reports the obfuscator as failed. Our correctness criteria and observational equivalence are defined in \S\ref{sec:equiv}.

\subsection{Sketch Definition}
\label{sec:sketch_def}
This section formally defines the sketch language used by OBsmith.
A sketch is a syntactically valid JavaScript program augmented with
placeholders that abstract over literals, variable references, and
expressions. Sketches are instantiated into concrete programs by a
separate sketch-filling procedure described in Section~\ref{sec:program-generation}.

\paragraph{Overview}

The sketch language is designed to satisfy two goals: (1) every sketch should be parsable as a JavaScript program by standard tooling (e.g., Babel), and
(2) a single sketch should compactly represent a large family of concrete
programs. To this end, placeholders are encoded using ordinary identifiers
and function calls, while semantic constraints such as scope resolution and
operator selection are enforced during sketch filling rather than in the
grammar.

\paragraph{EBNF Grammar}

Table~\ref{tab:sketch-ebnf} presents the EBNF grammar of the sketch language.
The grammar specifies only the syntactic structure of sketches; it does not
encode semantic constraints such as variable scope, type compatibility, or
randomized generation strategies.
\begin{table}[t]
\centering

\caption{EBNF grammar of sketch expressions embedded in JavaScript programs.}
\renewcommand{\arraystretch}{1.0}
\scalebox{0.9}{
\begin{tabular}{ll}
\hline
\textbf{Nonterminal} & \textbf{Production} \\
\hline
Program
  & $\{\, JavaScriptStatement \,\}$ \\

JavaScriptStatement
  & \textit{any JavaScript statement} \\

Expression
  & Atom \\
  & $\mid$ ArithmeticExpr \\
  & $\mid$ RelationExpr \\
  & $\mid$ LogicExpr \\

Atom
  & Identifier \\
  & $\mid$ Literal \\
  & $\mid$ Placeholder \\

Literal
  & NumberLiteral \\
  & $\mid$ BooleanLiteral \\

Placeholder
  & \texttt{numberLiteral} \\
  & $\mid$ \texttt{booleanLiteral} \\
  & $\mid$ \texttt{numberReference} \\
  & $\mid$ \texttt{booleanReference} \\

ArithmeticExpr
  & \texttt{arithmetic}\texttt{(}Expression\texttt{,} Expression\texttt{,} OperatorList\texttt{)} \\

RelationExpr
  & \texttt{relation}\texttt{(}Expression\texttt{,} Expression\texttt{,} OperatorList\texttt{)} \\

LogicExpr
  & \texttt{logic}\texttt{(}Expression\texttt{,} Expression\texttt{,} OperatorList\texttt{)} \\

OperatorList
  & Operator \\
  & $\mid$ Operator \texttt{,} OperatorList \\
\hline
\end{tabular}
}
\label{tab:sketch-ebnf}
\end{table}

\begin{table}[t]
\centering

\caption{Operator categories used in expression placeholders.}
\renewcommand{\arraystretch}{1.0}
\scalebox{0.9}{
\begin{tabular}{ll}
\hline
\textbf{Category} & \textbf{Operators} \\
\hline
ArithmeticOp
  & \texttt{+}, \texttt{-}, \texttt{*}, \texttt{/}, \ldots \\

RelationalOp
  & \texttt{<}, \texttt{>}, \texttt{<=}, \texttt{>=}, \texttt{==}, \texttt{===}, \ldots \\

LogicalOp
  & \texttt{\&\&}, \texttt{||} \\
\hline
\end{tabular}
}
\label{tab:operators}
\end{table}

\begin{table}[t]
\centering

\caption{Lexical categories used in the sketch grammar.}
\scalebox{0.9}{
\begin{tabular}{ll}
\hline
\textbf{Token} & \textbf{Description} \\
\hline
Identifier
  & JavaScript identifier \\

NumberLiteral
  & JavaScript numeric literal \\

BooleanLiteral
  & \texttt{true} $\mid$ \texttt{false} \\

OtherJSStatement
  & Any JavaScript statement not involving placeholders \\
\hline
\end{tabular}
}
\label{tab:lexical}
\end{table}

\paragraph{Operator Categories}

Operators appearing in expression placeholders are drawn from the categories
shown in Table~\ref{tab:operators}. Operator lists specify alternative
operators from which one is selected during sketch filling.

\paragraph{Lexical Categories}

Table~\ref{tab:lexical} summarizes the lexical categories referenced in the
grammar. These categories follow standard JavaScript syntax and are not
extended by the sketch language.

\paragraph{Discussion}

The EBNF grammar defines the syntactic shape of sketches and ensures that
every sketch is a valid JavaScript program. It intentionally omits semantic
constraints such as variable scope, operand compatibility, and operator
selection. These aspects are handled during sketch filling, where
placeholders are resolved and expression generators are instantiated into
concrete JavaScript expressions. This separation allows the sketch language
to remain simple while enabling flexible and diverse program generation.

\subsection{LLM-powered Sketch Generation}\label{sec:llm-sketching}

\begin{figure}[htbp]
    \centering
    \includegraphics[width=.6\linewidth]{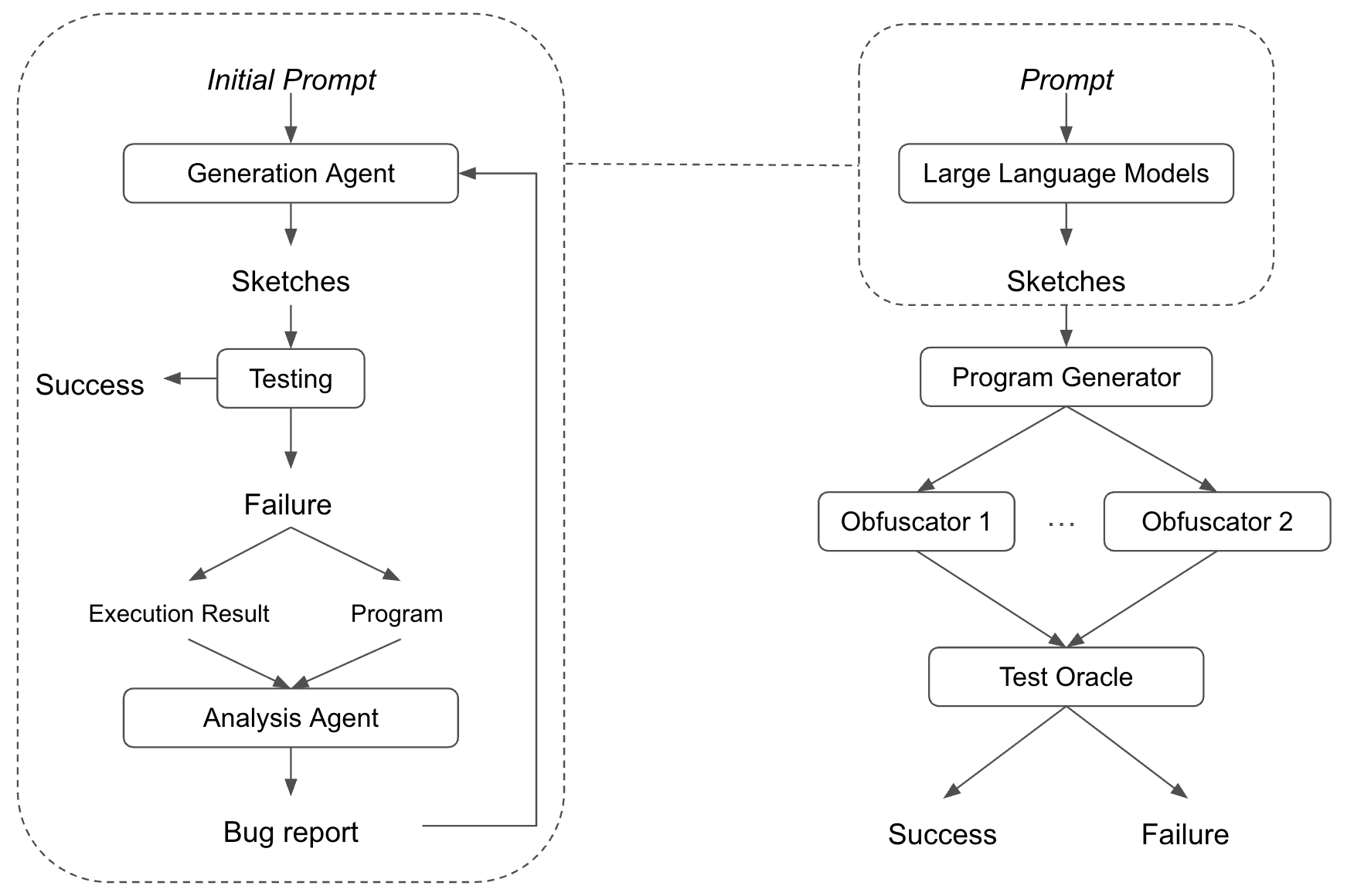}
    \caption{LLM-powered sketch generation with feedback loop (left) and corresponding location in OBsmith framework}
    \label{fig:approach}
\end{figure}
OBsmith generates \emph{sketches}—i.e., program templates with typed placeholders that capture domain knowledge about JavaScript—by prompting an LLM with a constrained template language and explicit formatting requirements (Fig. \ref{fig:approach}). The goal is to obtain (i) \emph{diverse} yet \emph{valid} templates that exercise representative control-flow and data-flow patterns, and (ii) sketches that can be deterministically instantiated by our program generator in \S\ref{sec:program-generation}. 

\paragraph{Motivation}
LLMs encode rich statistical priors over real-world code and idioms. By coupling a \emph{constrained} prompt with \emph{typed placeholders}, OBsmith harnesses these priors to synthesize compact, human-like templates that systematically expand into large, varied test suites—achieving breadth unattainable with purely random grammars while keeping generator complexity low. Empirically, this design yields high rates of syntactic validity and exposes correctness bugs across obfuscators and configurations (see \S\ref{sec:result}).

\paragraph{Prompt design.}
The prompt (Fig. \ref{fig:prompt} in appendix) specifies: (1) the sketching \textbf{DSL} (literal, reference and expression placeholders); (2) \textbf{usage constraints} (operands may be literals, references, or nested expressions); (3) an \textbf{I/O format} that yields exactly ten sketches in a fenced code block; and (4) a \textbf{worked example} that pairs a template with a filled instance to ground the LLM’s understanding of intended semantics. This mix of schema, constraints, and exemplars reduces drifting and encourages syntactically uniform outputs that downstream tooling can parse. 

\paragraph{Diversity controls.}

We ensure diversity at two stages: sketch synthesis and sketch instantiation.
(1) Sketch synthesis (LLM stage). Firstly, OBsmith's prompt explicitly asks for representative JavaScript idioms and control-flow constructs (e.g., loops, branches, function calls, and array operations) and requests distinct sketches per invocation, encouraging structural variety rather than repeated patterns. Furthermore, we generate sketches in a single LLM session, keeping prior outputs in context and explicitly prompting the model to avoid producing sketches similar to earlier ones.

(2) Sketch instantiation (filling stage). OBsmith then expands each sketch into many concrete programs by sampling at controlled variability points: literals are randomized; references are selected via scope analysis from in-scope variables of the appropriate type; and expression factories sample an operator from an explicit operator set (and may nest recursively to increase AST depth). Finally, we repeat this filling procedure multiple times per sketch, producing a set of distinct programs from the same template.

Together, these mechanisms yield diverse CFG/DFG shapes and expression structures from a compact sketch specification.

\paragraph{Validity guards in the prompt.}
Because program generation (\S\ref{sec:program-generation}) assumes sketches are \emph{parsable JavaScript with placeholders}, the prompt: (1) requires placeholders to appear where the resulting types are admissible (e.g., \texttt{numberLiteral} in numeric contexts, \texttt{booleanReference} in predicates); and (2) constrains expression factories to binary operators (logical, relational, arithmetic). These instructions align the LLM outputs with the OBsmith sketch syntax and reduce post-hoc repair.

\begin{figure}[htbp]
    \centering
    \includegraphics[width=.6\linewidth]{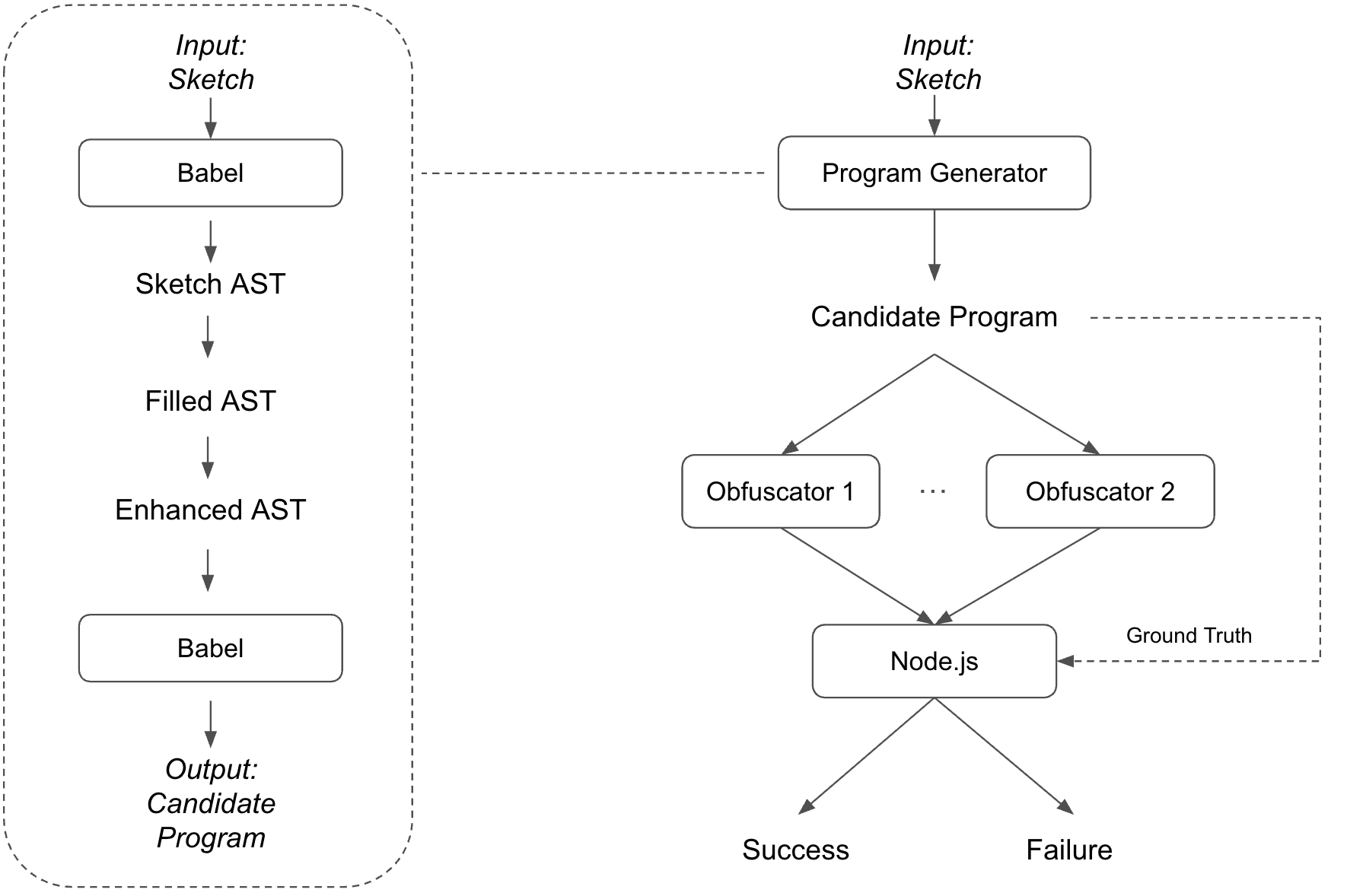}
    \caption{Program generation workflow (left) and reference-oriented equivalence testing workflow (right)}
    \label{fig:approach}
\end{figure}

\subsection{Feedback Loop for Sketch Generation}
\label{sec:feedback_loop}
To enhance the effectiveness of LLM-powered sketch generation, OBsmith incorporates a 
\emph{feedback loop} that automatically leverages prior test outcomes to guide the synthesis of new sketches. This mechanism transforms the testing process from a one-shot generation exercise into an adaptive, iterative cycle that continually improves coverage of bug-revealing scenarios.

\paragraph{Motivation.}
While LLMs can produce diverse and syntactically valid sketches from carefully designed 
prompts (\S\ref{sec:llm-sketching}), their output may drift toward common or uninformative patterns. At the 
same time, reference-oriented equivalence testing (\S\ref{sec:diff}) often produces valuable signals in the form of 
failures, inconsistencies, or execution traces that highlight semantic corner cases. The 
feedback loop is designed to exploit these signals: instead of discarding failures after 
bug discovery, OBsmith repurposes them to guide future sketch generation.

\paragraph{Agent Loop Architecture.}
The feedback loop of OBsmith employs a multi-agent framework with distinct roles:
\begin{enumerate}
    \item \textbf{Bug report analysis.} A LLM-based bug analysis agent is prompted to analyze the execution result and corresponding failing programs. It summarizes the discrepancy 
    (e.g., suppressed exceptions, scope violations, control-flow divergence) in the structured execution result. Then it generate a \emph{bug report} captures key features of the bug-triggering scenario while abstracting away irrelevant details.
    \item \textbf{Sketch refinement.} The second LLM takes the bug report as input and 
    generates new sketches that are likely to reproduce or generalize the observed issue. 
    For example, if the bug report highlights mishandling of \texttt{eval()} or 
    incorrect scoping of class constructors, the LLM is instructed to embed placeholders 
    involving these constructs into new sketches. 
\end{enumerate}

\paragraph{Interaction with follow up testing.}
The feedback loop works synergistically with reference-oriented equivalence testing (\S\ref{sec:diff}) and metamorphic testing (\S\ref{sec:metamorphic_testing}). While testing reveals reusable patterns from failing programs, the feedback loop ensures that LLMs actively explore related semantic neighborhoods. This combined mechanism balances data-driven generalization (via LLM) and deterministic reuse (via testing), enabling both novelty and stability in the evolving sketch corpus.

\paragraph{Benefits.}
By coupling bug-driven feedback with generative modeling, OBsmith can progressively 
concentrate its testing effort on areas where obfuscators are fragile. This adaptive 
process reduces redundant sketch generation, improves the precision of bug discovery, 
and ensures sustained effectiveness as new obfuscators or configurations are introduced. 
In practice, our evaluation (\S\ref{sec:result}) shows that the feedback loop substantially increases 
the number of distinct correctness bugs detected compared to generation without feedback.

\paragraph{Summary.}
The feedback loop elevates OBsmith from a static testing framework to a self-improving 
system. By turning runtime failures into structured prompts for sketch synthesis, it 
creates a virtuous cycle in which every discovered bug enriches the generator, leading 
to more robust and targeted testing in subsequent iterations.

\subsection{Sketch Extraction from Real-World Program}
\label{sec:sketch_extraction}
In addition to LLM-powered sketch generation (\S\ref{sec:llm-sketching}), another component of OBsmith's sketch generation is automatic sketch extraction, which aims to automatically extract reusable program sketches from real-world JavaScript programs. 

\paragraph{Motivation.}
Reference-oriented equivalence testing and metamorphic testing (\S\ref{sec:diff}, \S\ref{sec:metamorphic_testing}) frequently reveal 
program fragments that expose semantic inconsistencies in obfuscators. These fragments 
often encode subtle interactions between JavaScript features—such as scoping rules, 
exception propagation, or dynamic evaluation—that are difficult to anticipate in 
manual prompt design. Automatically converting such fragments into parameterized 
sketches increases test diversity while focusing future iterations on bug-revealing 
structures. Moreover, sketch extraction can automatically use existing JavaScript programs, which usually contain the developer's domain knowledge and have been proven to be effective for compiler testing \cite{lejit}. For the above two reasons, OBsmith also designed an automated framework to extract sketches from existing programs.

\paragraph{Extraction Pipeline.}
OBsmith implements the extraction at the AST level. It parses the concrete JavaScript program to a Babel AST and proceeds the sketch extraction in two steps:
\begin{enumerate}[label=Step \arabic*:]
    \item \textbf{Collect candidates.} Identify variables that hold numbers or booleans.
    
    \item \textbf{Replacement.} The program is parsed into an Abstract Syntax Tree (AST), and literals, variable references, and selected expressions are abstracted back into 
    OBsmith placeholders (\texttt{numberLiteral}, \texttt{booleanReference}, 
    \texttt{arithmetic()}, etc.). The generalization step ensures that the resulting 
    sketch is type-correct and can be re-instantiated with diverse values and expressions.
\end{enumerate}

\paragraph{Benefits.}
By automating the discovery of new sketches, OBsmith transforms one-off bug-triggering 
programs into systematic generators of test cases. This mechanism reduces manual effort, 
broadens semantic coverage, and enables continuous adaptation of the testing pipeline as 
new obfuscators or language features emerge. Importantly, sketch extraction allows OBsmith 
to evolve its corpus in tandem with the domain, turning transient failures into 
long-lasting assets for regression testing.

\paragraph{Summary.}
Together with LLM-powered sketch generation, program filling, and reference-oriented equivalence testing, 
sketch extraction completes a feedback-driven workflow that continually enriches OBsmith’s 
test suite. This synergy ensures that OBsmith is not only capable of finding bugs once, 
but also of systematically amplifying its effectiveness over time by reusing and refining 
bug-inducing patterns.

\subsection{Sketch-based Program Generation}
\label{sec:program-generation}
OBsmith generates sketches using LLMs and auto-extraction, and these sketches are input to the program generator. Program generation is a two-step process that transforms partial sketches into complete, executable JavaScript programs which are ready for equivalence checking.

\begin{enumerate}[label=Step \arabic*:]
    \item \textit{Sketch filling}. \name~systematically replaces all placeholders and incomplete expressions in each sketch. OBsmith traverses the sketch and substitutes each placeholder with a concrete value or expression, producing a filled concrete program.
    
    \item \textit{Program enhancing}. Filled programs are augmented with testing oracles that enable straightforward equivalence checking between original and obfuscated code. These oracles ensure that behavioral differences can be reliably observed during execution. The output of this stage is a set of candidate programs, ready to be obfuscated and evaluated.
\end{enumerate}

\subsubsection{Sketch filling}
\label{sec:sketch_fill}
The sketch filling is a specialized code transformation phase to transform an abstract sketch into multiple, unique, and concrete JavaScript programs. It systematically fills the literal and reference placeholders with random values or variables and uses an "expression generator" to fill expression placeholders within a sketch, effectively translating the high-level program template into executable JavaScript code. This process is the first and critical step in creating the diverse corpus of programs required for effective testing.

We implement sketch filling algorithm in Babel, which allows parsing with placeholders into an AST. For each program to be generated, the sketch is parsed into an AST using a standard Babel JavaScript parser. OBsmith traverses and fills placeholders using a visitor pattern. At each node, it checks for the presence of OBsmith's specialized placeholders. When a match is found, it performs a transformation. OBsmith's sketch filling logic is divided based on the type of placeholders encountered during the AST traversal.

\paragraph{Literal placeholders} (numberLiteral and booleanLiteral).
These are the simplest transformations, typically involving the replacement of an Identifier node in the AST. (numberLiteral, booleanLiteral): When an identifier matching one of these placeholders is found, it is replaced with a new Literal node in the AST. The value of this literal is a randomly generated primitive of the corresponding type. For booleanLiteral, OBsmith simply replaces it with a True or a False with equal probability. Since JavaScript only provides the Number class and doesn't distinguish between Integer and Float numbers, for numberLiteral, OBsmith generates a random integer or a random float with equal probability.

\paragraph{Reference placeholders} (numberReference and booleanReference). This is a more complex operation that requires scope analysis. When a reference placeholder is found, the script traverses up the AST from the current node to find all variable declarations in the current and parent scopes. It filters this list to find variables of the correct type and randomly selects one from the filtered results. The placeholder Identifier node is then replaced with a new Identifier node containing the name of the selected variable. If no suitable variable is found in current scope, OBsmith uses a corresponding boolean or number value to fill the sketch. 

It’s worth noting that in addition to using explicitly defined variables (\texttt{let a = 1; let b = True}), OBsmith also supports filling sketches with elements from arrays. Unlike other programming languages that contain strict type check mechanism, JavaScript allows elements of different types to exist in the same array. For example, \texttt{let arr = [True, 3.1415926, "Hello World!", 1];} is a valid statement in JavaScript. If \texttt{arr} is accessible in current scope, OBsmith can use arr[0] to fill booleanReference and use arr[1] or arr[3] to fill numberReference.

\paragraph{Expression placeholders.} (e.g. arithmetic(operand1, operand2, operator1, operator2 ...))
This is the most sophisticated part of the sketch filling algorithm. The expression placeholders are identified as CallExpression nodes in Babel AST. When a CallExpression node whose callee is one of the generators (arithmetic, relation, logic) is found, the script performs the following steps:

\begin{enumerate}[label=Step \arabic*:]
\item \textit{Argument Resolution:} The first two parameters represent the left and right operands, respectively. OBsmith recursively processes these two parameters passed to the generator function. If operand is another expression placeholder, the inner expression placeholder is parsed first. If it is a literal or reference placeholder, it is replaced as described above. This recursive population method correctly constructs deeply nested expression placeholders.

\item \textit{Operator Selection:} Starting with the third parameter, the operators that can be used for this expression placeholder are represented. OBsmith examines the operator parameter (which can be a string literal such as "+", "*", "===", or the operator itself +, *, ===). OBsmith randomly selects one of these operators to use in the final expression.

\item \textit{Node Transformation:} The original CallExpression node is removed from the AST and replaced with a new BinaryExpression node. This new node's left and right attributes (operand) are set to the parsed operand parameters, and its operator attribute is set to the randomly selected operator.
\end{enumerate}

\paragraph{Invalid expression syntax.} Although we have instructed LLMs to generate valid sketches through a series of prompting techniques as shown in \S\ref{sec:llm-sketching}, we still find that LLM generate syntactically incorrect sketches that are different from our intention. To enhance usability, OBsmith has designed error tolerance mechanisms for some common error types to reduce program generation failures caused by operator errors. Specifically, the following measures are taken for common operator errors.

\begin{enumerate}[label=Error \arabic*:]
\item \textit{Use unary operator in expression placeholder.} If a unary operator $unary\_op$ appears in an expression placeholder and $unary\_op$ is selected when filling sketch, OBsmith generates a unary expression node using the first operand and ignoring the second operand. The unary expression node will replace the original expression placeholder.
    
\item \textit{No operator in expression placeholder.} If there is no operator in the expression placeholder, and only two parameters represent the left and right operands, OBsmith will randomly select one from all Binary operators and use these two parameters as the left and right operands to generate a binary expression node, which will replace the original expression placeholder.
\end{enumerate}

After the AST traversal and filling are complete for one iteration, the modified AST—which no longer contains any OBsmith defined placeholders—is passed to Babel, which converts the AST back into a syntactically correct JavaScript program. The process is repeated multiple times to produce a set of distinct programs from a single sketch, with randomness introduced at specific, controlled placeholders.

\subsubsection{Program Enhancing}\label{sec:enhance}
Program enhancement is a source-to-source transformation that takes a filled JavaScript program (in \S\ref{sec:sketch_fill})as input and systematically injects logging code to produce an enhanced program. The goal is to make the program's execution process observable and deterministic but not change its behavior and functionality. OBsmith implements the program enhancement in Babel. The input (filled) program is first parsed into an AST, which OBsmith traverses top-down using a visitor pattern. During traversal, new \texttt{ExpressionStatement} nodes are injected at locations such as \texttt{BlockStatement} or \texttt{Function\-Declaration}, with each \texttt{ExpressionStatement} node containing a \texttt{CallExpression} (e.g., \texttt{console.log}) for logging. After traversal, the enhanced AST is converted back to JavaScript code and written to an output file. The enhancement involves several different types of logging code injection, each designed to provide different aspects of runtime observability. Specifically, we inject the following logging instruments.

\paragraph{Global error handling.} JavaScript is a lightweight, function-first, interpreted (or just-in-time compiled) language that executes code line by line. Because of this characteristic, JavaScript lacks a main function, making it difficult to determine the complete execution of a program. Therefore, we designed global error handling to wrap the entire program body in a global try...catch block. Specifically, the program's top-level statements are moved into the body of a TryStatement node. A CatchClause is added to handle any exceptions, logging the caught error and the error message. This ensures that any uncaught exceptions during program execution are caught and logged in a standard format (e.g., "!!! GLOBAL ERROR Caught!!!"). When an error is caught, this mechanism also forces the program to exit with a non-zero status code (process.exit(1)), clearly signaling failure to the test script.

\paragraph{Block level control flow tracing.} OBsmith identifies every block statement in the AST. A block statement is a sequence of statements enclosed in curly braces, such as a function body in function declaration node, the consequent and alternate block of an if statement, or the body of a for or while loop. OBsmith contains both block entry and block exiting log. Specifically, a console.log statement is prepended to the beginning of every block. This log indicates entry into the block, using its location in the source file as a unique identifier (e.g., "-> Entering Block@<line>:<col>"). A console.log statement is appended to the end of every block to signal its completion (e.g., "<- Exiting Block@<line>:<col>").
 
\paragraph{Data flow tracing via state checksum.} This is a critical instrumentation for detecting data-flow bugs. Just before the exit log of every block, another console.log is injected. This statement performs scope analysis to identify all  accessible variables from current block. It then logs the name and current value of each variable. The Log Format is as follows " --- Checksum for Block@<line>:<col> --- var1: value1, var2: value2, ..." This "checksum" provides a snapshot of the program's data state at the end of every basic block. OBsmith uses this output to verify that the data state of an obfuscated program matches the baseline at every step of execution.

\paragraph{Function call instruments.} To observe the arguments passed to functions, additional logging is injected. At the beginning of every function's body (immediately after the block entry log), a console.log is added. The Log Format is "=> Entering function: <functionName> Arguments: <arg1>, <arg2>, ..." This log makes the dynamic call graph and the data passed between functions explicit, allowing for the detection of bugs where an obfuscator might reorder or incorrectly modify function arguments.

\subsection{Correctness Criteria and Observational Equivalence}
\label{sec:equiv}

\paragraph{Goal and scope.}
An obfuscator is intended to be \emph{semantics-preserving}: the obfuscated program should exhibit the same behavior as the original one when executed in the same runtime environment (e.g., a fixed \texttt{Node.js} version). Because fully proving semantic equivalence for JavaScript is infeasible in general, OBsmith adopts an \emph{observational} equivalence criterion based on a normalized \emph{observable trace} that is (i) substantially stronger than comparing only final outputs, and (ii) directly checkable at scale.

\paragraph{Observable trace.}
Given a program $P$ and an input $x$ (OBsmith programs may take no explicit input; $x$ may be empty), we execute $P$ under a fixed environment $E$ with OBsmith instrumentation enabled and obtain a sequence of runtime events:
\[
  \mathsf{Trace}_E(P, x) = \langle e_1, e_2, \dots, e_n \rangle.
\]
Each event $e_i$ belongs to a finite set of observable event kinds, including:
\begin{itemize}
  \item \textbf{I/O events:} $\mathsf{Out}(s)$ and $\mathsf{Err}(s)$ for normalized writes to \texttt{stdout}/\texttt{stderr}.
  \item \textbf{Termination events:} $\mathsf{Exit}(k)$ for a normal exit code $k$, and $\mathsf{Timeout}$ if the run exceeds the time budget.
  \item \textbf{Exception events:} $\mathsf{Throw}(\tau, m)$ for uncaught exceptions, recording (a normalized form of) the exception type $\tau$ and message $m$.
  \item \textbf{Control-flow events:} $\mathsf{Enter}(b)$ and $\mathsf{Leave}(b)$ when entering/leaving an instrumented basic block $b$.
  \item \textbf{Data-flow summaries:} $\mathsf{State}(b, h)$ at selected boundaries, where $h$ is a checksum of in-scope variable values at block $b$.
  \item \textbf{Call events:} $\mathsf{Call}(f, \vec{a})$ recording a normalized callee identifier $f$ and argument summary $\vec{a}$ for selected calls.
\end{itemize}

\paragraph{Normalization.}
To make traces comparable across runs, OBsmith applies a deterministic normalization function $\mathsf{Norm}(\cdot)$ to event payloads (e.g., canonicalizing line endings, eliding runtime-specific prefixes in stack messages, and serializing values into a stable representation before hashing). In particular, $\mathsf{State}(b,h)$ uses a stable serializer and a hash function over primitive values and bounded summaries of objects/arrays to avoid brittleness due to formatting differences while still detecting semantic drift. We denote the resulting normalized trace by:
\[
  \widehat{\mathsf{Trace}}_E(P, x) \triangleq \mathsf{Norm}\!\left(\mathsf{Trace}_E(P, x)\right).
\]

\paragraph{Operational semantic equivalence.}
We define two programs $P$ and $Q$ to be \emph{observationally equivalent} under environment $E$ and input $x$, written $P \approx_{E,x} Q$, iff their normalized traces are identical:
\[
  P \approx_{E,x} Q \quad \Longleftrightarrow \quad
  \widehat{\mathsf{Trace}}_E(P, x) = \widehat{\mathsf{Trace}}_E(Q, x).
\]
This relation is intentionally stronger than ``same final output'': it requires agreement on termination mode (normal exit vs.\ exception vs.\ timeout), the dynamic control-flow path (block enter/leave sequence), and a coarse-grained but effective approximation of data-flow and call behavior (state checksums and call-argument summaries).

\paragraph{Correctness criterion for obfuscation.}
Let $\mathcal{O}$ be an obfuscator and $c$ an obfuscation configuration. OBsmith tests correctness by executing the original program $P$ and its obfuscated version $\mathcal{O}_c(P)$ on the same input $x$ and checking:
\[
  P \approx_{E,x} \mathcal{O}_c(P).
\]
Any violation indicates a semantic divergence \emph{in the tested environment} $E$ and under the operational criterion above.

\paragraph{Handling nondeterminism.}
JavaScript programs and runtimes can exhibit nondeterminism (e.g., due to time, randomized hashing, or scheduling). OBsmith therefore (i) avoids nondeterministic APIs in generated tests by construction when possible, and (ii) reruns executions multiple times: a discrepancy is reported as a bug only if it reproduces consistently across reruns of both $P$ and $\mathcal{O}_c(P)$.

\paragraph{Limitations.}
Our criterion does not claim full semantic equivalence for all JavaScript behaviors, especially those involving external I/O, the network, or environment-dependent APIs. Instead, it provides a practical, scalable, and substantially stronger-than-output notion of equivalence that is well-suited for detecting semantic drift introduced by source-to-source obfuscation.

\subsection{Reference-Oriented Equivalence Testing}
\label{sec:diff}

Differential testing is a widely adopted technique in compiler validation~\cite{csmith,jattack}, 
where the absence of a reliable ground truth necessitates comparing the outputs of multiple 
compiler variants on the same input program. In contrast, OBsmith operates under the assumption that the original JavaScript program, executed directly in a trusted engine (Node.js), serves as the reference oracle (ground truth). This allows OBsmith to adapt  reference-oriented equivalence testing  specifically for the obfuscation domain.

\paragraph{Methodology.} 
Given a candidate program generated from a sketch, OBsmith applies multiple obfuscators (e.g., 
JS-Confuser, Obfuscator.IO) under different configuration settings. The original and obfuscated 
programs are then executed within Node.js. OBsmith collects and normalizes their execution 
outputs, where the output of the original program is treated as the reference oracle.

\paragraph{Equivalence Checking.} 
For each obfuscated variant, OBsmith compares its execution results with the oracle. The 
comparison includes:
\begin{itemize}
    \item \textbf{Standard Output:} Console outputs must be identical across all runs. 
    \item \textbf{Exceptions:} Exception types and messages must match; stack traces are ignored 
          due to nondeterministic file names and line numbers.
    \item \textbf{Termination Behavior:} Programs must terminate consistently. Divergences 
          such as premature exits or runtime crashes are flagged as failures.
\end{itemize}
A mismatch in any category constitutes a correctness bug introduced by the obfuscator.

\paragraph{Timeout Handling.}
Randomly generated candidate programs may introduce non-terminating behavior 
(e.g., \texttt{while(true)}). To avoid indefinite execution, OBsmith enforces a strict timeout of 
60 seconds. If both the original and obfuscated programs exceed this limit, the test case is 
considered consistent and marked as a pass. Otherwise, the discrepancy is recorded as a failure.

\paragraph{Summary.}
By leveraging the original program as a reference oracle, OBsmith strengthens differential testing with a reliable ground truth, termed reference-oriented equivalence testing. This adaptation enables precise detection of semantic inconsistencies introduced by obfuscation, distinguishing it from traditional approaches that rely on consensus among potentially flawed variants.

\subsection{Metamorphic Testing}
\label{sec:metamorphic_testing}
Reference-oriented equivalence testing  (\S\ref{sec:diff}) is our primary mechanism for checking semantic
equivalence between an original program and its obfuscated counterpart. In addition,
we explore whether \emph{metamorphic testing} can stress obfuscators by applying
equivalence-preserving rewrites to the \emph{inputs} before obfuscation and then
verifying that obfuscation does not change program behavior. The goal here is not
limited to the “single-obfuscator available” setting; rather, it is to probe whether
obfuscators are robust to semantics-preserving variation in their inputs.

\paragraph{Motivation.}
Many source-level rewrites leave a program’s observable semantics unchanged. If an
obfuscator is semantics-preserving, then applying such a rewrite to a program \emph{before}
obfuscation should not alter the behavior of the resulting obfuscated code. For example, constant folding, algebraic rewriting, or renaming of 
variables should not alter the runtime behavior of a program. If applying such a 
semantics-preserving transformation before and after obfuscation produces divergent 
outputs, the obfuscator has introduced a correctness bug.

\paragraph{Metamorphic Relations.}
OBsmith defines a suite of metamorphic relations (MRs) over JavaScript programs. Our implementation use three most frequently used MRs in compiler testing:
\begin{itemize}
    \item \textbf{Algebraic equivalence:} Replacing $x + 0$ with $x$, $x \times 1$ with $x$, 
    or $(x + y) - y$ with $x$.
    \item \textbf{Control-flow equivalence:} Transforming \texttt{if (true) \{S\}} into $S$, 
    or flattening nested conditionals into equivalent disjunctions.
    \item \textbf{Dead-code injection/removal:} Adding or removing statements that provably 
    do not affect program outputs (e.g., \texttt{if (false) \{...\}}).
\end{itemize}
Each MR captures a known semantic invariant. The transformations are automatically applied 
to programs generated from sketches, and both the original and transformed variants (by MRs) are processed by the obfuscator under test.

\paragraph{Testing Workflow.}
For each sketch instantiation:
\begin{enumerate}
    \item Generate a concrete program $P$ and one or more metamorphic variants 
    $\{P_1, P_2, \dots\}$ using the MRs above.
    \item For each obfuscator $O$ under test, apply $O$ with different configuration 
    settings to $P$ and to each $P_i$, producing multiple obfuscated versions 
    $\{O(P), O(P_1), \dots\}$.
    \item Execute all obfuscated programs under the same runtime environment, 
    collecting their outputs and observable side effects.
    \item Report a correctness bug if any obfuscated version of $P$ and its 
    corresponding $P_i$ differ in output, runtime exception behavior, or 
    termination status.
\end{enumerate}

\paragraph{Benefits.}
Metamorphic testing directly evaluates an obfuscator’s \emph{invariance} to
semantics-preserving rewrites, exposing order-sensitivity and brittle interactions
between transformation passes (e.g., encoder/decoder pipelines, control-flow
flatteners, scope virtualizers). It is equally applicable when multiple obfuscators
are available and when only one is present, and it supports lightweight regression
tests across versions/configurations of the same tool.

\paragraph{Relations to obfuscation.}
Metamorphic relations (MRs) provide \emph{test amplification} rather than defining correctness. For an MR $m$, OBsmith first checks that $m$ preserves behavior on the original program, i.e., $P \approx_{E,x} m(P)$. Only then does it check that obfuscation preserves this equivalence as well, by comparing $\mathcal{O}_c(P)$ and $\mathcal{O}_c(m(P))$ against their respective references. This separates the \emph{system under test} (the obfuscator) from the \emph{controlled transformations} used to generate additional, independently-checkable test inputs.

\paragraph{Summary.}
By encoding program equivalences as metamorphic relations, OBsmith provides
a lightweight yet powerful mechanism to validate obfuscators in the absence of oracles. This
technique, in combination with reference-oriented equivalence testing, yields comprehensive correctness checking across both multi-obfuscator and single-obfuscator scenarios

\section{Results}\label{sec:result}
To evaluate OBsmith, we study the following research questions:
\begin{itemize}
\item \textbf{RQ1:} How well LLMs follows the prompt and generate syntactically valid sketches?
\item \textbf{RQ2:} What critical bugs does OBsmith detect in real-world JavaScript obfuscators?
\item \textbf{RQ3:} How effective is OBsmith compared with the state-of-the-art techniques? 
\item \textbf{RQ4:} What are the contributions of the major components of OBsmith?
\end{itemize}

\subsection{Experimental Setup}
\subsubsection{Large language models}
Regarding LLM selection, we used the most capable LLMs available at the start of the experiment. For each LLM, we generate 100 different sketches. We also test the coding agent and see its effectiveness in generating sketchs compared to directly use base model. For fair comparison since coding agent like Gemini-CLI or Claude Code can see the whole project, we generate sketches first and then put other LLMs results into our project directory. Specifically, we use the following models: Gemini 2.5 Pro, GPT-5, Claude 4 Opus, Qwen3-235B-A22B-2507, Grok 4. Besides, we also evaluate LLM agent's ability in sketch generation. We use Gemini CLI with Gemini 2.5 Pro to generate 100 sketches as well. In order to fair comparison, we begin with Gemini CLI without other models' results, in other words, we do this first.

\subsubsection{Obfuscators under test}
\label{sec:obfuscation}
Based on Google’s investigation \cite{jiang2025cascade}, Obfuscator.IO \cite{obio} and JS-Confuser \cite{jsconfuser} are among the most widely used JavaScript obfuscators in industrial production. We therefore evaluate on these two tools to study failures in widely deployed obfuscation pipelines. We focus on two obfuscators to enable systematic coverage across configurations and to support thorough triage and root-cause analysis, rather than spreading the evaluation budget thinly across many tools.

Both tools take as input a candidate program and a JSON configuration that specifies obfuscation options. Because an obfuscator’s behavior can vary substantially with its settings, we run each candidate program under multiple configurations to generate a diverse set of obfuscated variants and increase coverage of the transformation space.

JS-Confuser provides three preset configurations (Low, Medium, High), and we use them without modification. Obfuscator.IO provides four presets (Default, Low, Medium, High), which we also use, with a small number of necessary adjustments for compatibility with our pipeline. Our implementation of OBsmith checks program equivalence using console.log; however, Obfuscator.IO’s Low/Medium/High presets disable the console, so we set the corresponding option to false. In addition, Obfuscator.IO’s High preset enables debug protection, which can cause scripts to hang forever, so we disable that option as well. Finally, we disable compact mode (which places the entire program on a single line), because V8 may reject extremely long lines with an “invalid size” error. We make no other configuration changes.


\subsubsection{JavaScript Engine}
We use Node.js to run our experiments. At its core, Node.js is a server-side JavaScript runtime environment built on the foundation of the V8 JavaScript engine. V8 is developed by Google and used in its Chrome browser, which is responsible for taking the JavaScript code written in a Node.js application and compiling it into the machine code that the computer can execute directly. This compilation process, leveraging V8's Just-In-Time (JIT) compilation and optimization techniques, allows Node.js applications to achieve high levels of performance and efficiency, particularly in handling concurrent connections and asynchronous operations. Essentially, V8 provides the raw power of JavaScript execution, while Node.js extends that power with a rich set of APIs and modules, making it a complete environment for building scalable server-side applications.

\subsubsection{Extraction dataset}
To evaluate sketch extraction, we uniformly sampled 1{,}000 source files  from the 150k JavaScript Dataset~\cite{data150k}. For each file, our extractor produced a single sketch. We then instantiated every sketch into three concrete programs, yielding a total of 3{,}000 instantiated programs for downstream evaluation.

\subsubsection{Baseline}
We select five JavaScript fuzzers (FuzzJIT~\cite{fuzzjit}, Jsfunfuzz~\cite{jsfunfuzz}, DIE~\cite{park:die}, Fuzzilli~\cite{gross2018fuzzil}, and Superion~\cite{wang2019superion}) as baselines. FuzzJIT~\cite{fuzzjit} is an oracle-enhanced fuzzing technique to detect non-crashing and crashing JIT compiler bugs. Jsfunfuzz is a generation based fuzzer for JavaScript/Node packages (practical for fuzzing JS libraries). DIE advances the crossover by restricting the types of sub-tree. Fuzzilli is a coverage-guided fuzzer for
JavaScript engines based on a custom intermediate language,
FuzzIL, which can be mutated and translated to JavaScript.
Instead of mutating the AST, or other syntactic elements of
a program, FuzzIL facilitates convenient mutations on the
control and data flow of a program. A FuzzIL program contains a list of instructions, and can be lifted to a JavaScript program for testing. Fuzzilli is popularly adapted to
build powerful JavaScript fuzzer by academia researchers and
industry practitioners. Superion finds bugs by conducting crossover on the AST sub-trees of two parent samples.

In our comparison experiments, we use the implementations of baselines except FuzzJIT provided by UniFuzz~\cite{unifuzz-li}, which is a fuzzing approach evaluation benchmark and provides a collection of docker for 37 well-known fuzzers, including Superion, DIE, Jsfunfuzz, and Fuzzilli, to ease the evaluation of different fuzzing approaches. Only Superion and DIE require initial
seeds. DIE’s initial seed corpus contains 100 JavaScript files
generated by its given script, and the same set of initial seeds
are also used for Superion. For FuzzJIT, we use their official implementation.

\subsubsection{Test budget.}
We budget testing effort by the number of generated candidate programs (i.e., the number of inputs submitted to the obfuscator), rather than by wall-clock time. This choice is motivated by the high variability of LLM generation latency, which can differ substantially across runs and prompts and becomes even less predictable when enabling \emph{thinking}-style inference. A time budget would therefore conflate testing effectiveness with model-side latency and infrastructure effects, making comparisons between LLM-based generation and traditional fuzzers inherently unfair.
Moreover, the two approaches differ fundamentally in their search strategy and computational profile: LLM-based synthesis is driven by token-level inference (sensitive to prompt length, retained context, and inference mode), whereas traditional fuzzers typically explore the input space via lightweight mutations guided by coverage feedback and are primarily bounded by execution throughput. Under a wall-clock budget, these mismatched dynamics would bias results toward whichever approach happens to be cheaper under a particular model and hardware configuration. By contrast, a test-count budget isolates effectiveness per generated input and yields more reproducible comparisons.

\subsection{RQ1: LLMs' Sketch Generation Ability}
\begin{table}[htbp]
\centering
\caption{Syntax validity and error types in LLM-generated sketches, each model generate 100 sketches}
\scalebox{0.95}{
\begin{tabular}{lcccccc}
\hline
\toprule
       & Valid Syntax & No op & Ternary op & Unary op & placeholder & Self-defined op\\
\midrule
Gemini 2.5 Pro & 92 & $\times$     &         &   $\times$    & &  $\times$   \\
GPT 5   & 98 &      &          &     $\times$   \\
Claude 4 & 96 &       &    $\times$      &    $\times$       \\
Qwen 3   &   88     &   $\times$    & $\times$ & $\times$ &      &         \\
Grok 4   &    69   &    $\times$   & & $\times$ & $\times$ & $\times$        \\
Gemini CLI    & 94 &   $\times$   &    &  $\times$  &     &    \\
\bottomrule
\end{tabular}
}
\label{table:sketch}
\end{table}

Table \ref{table:sketch} summarizes the quality of sketches produced by different LLMs. We evaluate (i) syntactic validity of generated sketches and (ii) adherence to the sketch spec, focusing on five recurring error modes: No-op (missing operator in a composite expression, especially in nested expressions), Ternary op (using ternary where our DSL forbids it), Unary op (introducing unary ops in expressions), Placeholder (using unsupported placeholder types), and Self-defined op (invented operators not in our grammar).

\paragraph{Overall performance.} The strongest models—GPT-5 (98\% valid), Claude 4 (96\%), and Gemini 2.5 Pro (94\%)—largely followed the prompt and DSL. Their residual errors were rare and concentrated in operator handling. Our error-tolerance pass converts sketches with a missing operator or stray unary operator into runnable programs, preventing pipeline crashes. \emph{Ternary op}, \emph{Unsupported Placeholder}, and \emph{Self-defined op} violations are unrecoverable within our DSL.

\paragraph{Per-model breakdown.}
\begin{itemize}
\item \textbf{GPT-5 (98\% valid).} Almost fully compliant; we observed no systematic category of spec violation in the sampled set. Occasional minor formatting irregularities were corrected by the sanitizer.
\item \textbf{Claude 4 (96\% valid).} High compliance overall, with a small number of \emph{ternary} insertions (e.g., embedding \texttt{cond ? a : b} inside a placeholder). These violate our sketch grammar.
\item \textbf{Gemini 2.5 Pro (92\% valid).} Solid adherence with two recurring operator issues: (i) \emph{No-op}—omitting a logical or arithmetic operator inside a nested expression; and (ii) \emph{Unary op}—placing \texttt{!} or \texttt{++} inside placeholders. Both are handled by our tolerance mechanism (assigning a safe default operator; stripping/hoisting the unary op).
\item \textbf{Qwen 3 (88\% valid).} Partial task understanding but frequent violations of our constraints. The dominant errors are \emph{No-op} omissions and \emph{Ternary op} use inside placeholders, which cannot be auto-repaired and would otherwise crash sketch filling.
\item \textbf{Grok 4 (69\% valid).} Most non-compliant. In addition to \emph{No-op} and \emph{Unary op} misuse, Grok frequently introduced \emph{Placeholder} types outside our DSL (strings) and Grok is the only model to fail in this area. Same as Qwen, Grok also \emph{Self-defined op} which is out of our sketch syntax.
\item \textbf{Gemini CLI (94\% valid).} Using the same base model as Gemini 2.5 Pro but with repository context, the agent followed the sketch–fill–enhance workflow reliably. The main residual issue was occasional \emph{Unary op} and \emph{No-op} omissions.
\end{itemize}

\paragraph{Code-agent vs. base model.} The Gemini CLI agent (94\% valid) shows that lightweight code awareness—file-system context plus access to repository scripts and documentation—substantially reduces spec violations relative to its base model. Residual issues are mostly occasional missing-operator (“no-op”) cases. Importantly, all Gemini CLI errors are automatically recoverable by our preprocessing/filling scripts, likely because the agent adheres more closely to the sketch–fill–enhance interfaces exposed by the repository.

\begin{center}
\setlength{\fboxrule}{1pt} 
\setlength{\fboxsep}{6pt} 
\fcolorbox{black!80}{gray!10}{
  \parbox{\dimexpr\columnwidth-2\fboxsep-2\fboxrule\relax}{%
    Answer to RQ1: GPT-5, Claude 4, and Gemini 2.5 Pro generate sketches that are both syntactically valid and DSL-conformant in the vast majority of cases, with residual issues concentrated in operator specification that our tolerance pass can absorb. Qwen 3 and Grok 4 struggle with domain-specific constraints, producing unrecoverable DSL violations. Providing task context via a code agent improves adherence without changing the underlying LLM.
  }%
}
\end{center}


\subsection{RQ2: Case Study}

Using OBsmith, we systematically reveal correctness bugs in both Obfuscator.IO and JS-Confuser. All identified bugs were reproducible and confirmed across both the online version in the obfuscator's website and the latest releases available in their respective GitHub repositories.  

\subsubsection{Bugs in JS-Confuser}
\begin{itemize}
    \item \textbf{Constructor name corruption}: JS-Confuser alters the constructor name of classes. For example, evaluating \texttt{console.log(bx.constructor.name)} produces incorrect results across all three configuration levels.  
    \item \textbf{eval() crashes}: Programs that execute correctly in their original form crash after obfuscation, consistently across all configurations.  
    \item \textbf{Silent ``fixes'' of exceptions}: In medium and high configurations, JS-Confuser modifies behavior so that programs that should raise exceptions instead complete execution without error.  
    \item \textbf{Incorrect error handling}: In medium and high configurations, expected exceptions are either suppressed or altered, resulting in the loss of valuable debugging information.  
    \item \textbf{Scope violations}: Variables accessed out of scope return \texttt{undefined} post-obfuscation, whereas debug information was available beforehand.  
    \item \textbf{Control-flow modifications}: In some medium-level configurations, the control flow is altered, leading to different return values. For instance, programs that should terminate with an exception instead end normally.  
    \item \textbf{Undefined value mishandling}: Similar to the control-flow issues, \texttt{undefined} values are silently ``corrected,'' altering return types. This occurs in medium and high configurations, effectively masking underlying errors.  
\end{itemize}

\subsubsection{Bugs in Obfuscator.IO}
\begin{itemize}
    \item \textbf{Unhandled constructs}: Certain valid JavaScript constructs cannot be obfuscated. Examples include:
    \begin{itemize}
        \item \texttt{async function () \{ \} / 1;}
        \item \texttt{class x \{ static \{ function x() \{ \} function x() \{ \} \} \}}
        \item \texttt{! class \{ \}();}
    \end{itemize}
    \item \textbf{Type errors}: Some transformed programs result in runtime \texttt{TypeError}s, breaking semantic equivalence with the original code.  
\end{itemize}

\subsubsection{Error-prone obfuscation techniques.}
Our failures exhibit clear configuration dependence, but the dominant risk factors are as follows. 
For \textsc{JS-Confuser}, we observed that more aggressive presets (e.g., Medium/High) are more likely to trigger semantic deviations because they enable transformations that interact poorly with JavaScript’s dynamic semantics and reflective features. A recurring error-prone family concerns \emph{exception and unresolvable-identifier handling}: obfuscation may replace a required \texttt{ReferenceError} with the value \texttt{undefined}, which can change control flow (e.g., altering which \texttt{try/catch} paths execute) or silently propagate \texttt{undefined} into program state. A second family involves \emph{renaming under reflection}: aggressive renaming can break \texttt{eval}-based code when string-evaluated expressions refer to original identifiers, and it can change observable identifier metadata such as \texttt{constructor.name}. These observations suggest that JS-Confuser configurations enabling aggressive renaming and exception-related rewrites are comparatively riskier, while less aggressive presets are safer when applications rely on precise exception behavior or reflection.

For \textsc{Obfuscator.IO}, the main configuration-dependent issue we observed is associated with \emph{control-flow flattening}, which can lead to a runtime \texttt{TypeError}. In contrast, the other crashes we encountered are \emph{parsing-stage failures} that persist across all tested presets, suggesting that they are less sensitive to configuration choices. We emphasize that these findings are empirical observations from our test suite rather than universal guarantees.

\begin{center}
\setlength{\fboxrule}{1pt} 
\setlength{\fboxsep}{6pt} 
\fcolorbox{black!80}{gray!10}{
  \parbox{\dimexpr\columnwidth-2\fboxsep-2\fboxrule\relax}{%
    Answer to RQ2: These findings demonstrate that correctness bugs are pervasive in widely used obfuscators. While some errors manifest as program crashes, others are more insidious---such as silent changes in return values or suppressed exceptions---that can undermine program reliability and complicate debugging. OBsmith’s LLM-driven, sketch-based testing proves highly effective at surfacing such issues, reinforcing the need for rigorous correctness validation in obfuscator development.
  }%
}
\end{center}

\subsection{RQ3: Comparison Experiment}

\begin{table*}[t]
\centering
\caption{Comparison with baseline and ablation study, Issue IDs refer to GitHub issues in OWNER/REPO.}
\renewcommand{\arraystretch}{1.3}
\scalebox{0.85}{
\begin{tabular}{l cccccc ccccc}
\toprule
\multirow{2}{*}{\textbf{Bug [Issue ID]}} &
\multicolumn{6}{c}{\textbf{Comparison with baselines}} &
\multicolumn{5}{c}{\textbf{Ablation study}} \\
\cmidrule(lr){2-7} \cmidrule(lr){8-12}
 & OBsmith & FuzzJIT & Jsfunfuzz & Superion & DIE & Fuzzilli 
 & -l & -f & -e & -m & -r \\
\midrule
constructor name [185] & \checkmark & \checkmark & \checkmark & \checkmark &  \checkmark & \checkmark & \checkmark &\checkmark & \checkmark& & \checkmark \\
eval crash [186] & \checkmark & & & & & & \checkmark & & & & \checkmark \\
silent fix [188] & \checkmark & & & & & & \checkmark& & & & \checkmark \\
error handling [188] & \checkmark & & & & & & \checkmark& \checkmark& \checkmark& & \checkmark \\
scope violation [189] & \checkmark & & & & & & \checkmark& & \checkmark& & \checkmark \\
control flow change [187] & \checkmark & \checkmark & \checkmark & & & & \checkmark& \checkmark& \checkmark& & \checkmark \\
undefined handling [189] & \checkmark & & & & & & \checkmark& \checkmark& & & \checkmark \\
async crash [1354] & \checkmark & &\checkmark & & & & & &\checkmark & & \checkmark \\
class crash [1354] & \checkmark & & & & & & \checkmark& \checkmark& & & \checkmark \\
! class crash [1354] & \checkmark & & & & & & \checkmark& & & & \checkmark \\
type errors [1289] & \checkmark & & & & & & \checkmark & & & & \checkmark \\
\bottomrule
\end{tabular}
}
\label{tab:compare}
\end{table*}

We compare \textsc{OBsmith} against five JavaScript fuzzers (FuzzJIT, JSFuzz, Superion, DIE, and Fuzzilli) on the same backend (V8 as shipped in Node.js) under an equal program budget. Concretely, we use six generators—five LLMs plus one agent (Gemini CLI) —to produce \textbf{600} sketches (100 per LLMs) and instantiate \textbf{5} programs per sketch, yielding \textbf{3{,}000} programs. Each technique then exercises V8 on this budget, and we record whether a technique exposes a unique bug (newly observed misbehavior) under our oracle. Baselines were taken from UniFuzz where available; Superion and DIE were supplied with the required default seed inputs.

\paragraph{Results.} Table~\ref{tab:compare} shows that \textsc{OBsmith} is the \emph{only} technique that exposed the bugs we report under this setting; none of the five fuzzers reproduced these failures within the same execution budget. 
The correctness regressions we surface (e.g., exception suppression, scope violations, constructor-name corruption, and eval() crashes) were all triggered by programs produced via our sketch–fill–enhance pipeline and were \emph{not} rediscovered by general-purpose engine fuzzers in our runs. 

\paragraph{Interpretation.} This outcome is expected: the baselines are optimized to find \emph{engine} defects (JIT/VM bugs) via coverage-guided mutation, while \textsc{OBsmith} is specialized for validating \emph{semantics-preserving transformations} by obfuscators, with an oracle that checks output/exception/termination equivalence across original and transformed programs. The comparison therefore demonstrates complementary strengths rather than a head-to-head replacement.

\begin{center}
\setlength{\fboxrule}{1pt} 
\setlength{\fboxsep}{6pt} 
\fcolorbox{black!80}{gray!10}{
  \parbox{\dimexpr\columnwidth-2\fboxsep-2\fboxrule\relax}{%
   Answer to RQ3: Under an equal program budget (3,000 programs), \textsc{OBsmith} surfaced all observed correctness bugs, whereas state-of-the-art JavaScript fuzzers did not expose these issues in our setting. The evidence supports using \textsc{OBsmith} alongside engine fuzzers: the former targets transformation-induced semantic drift; the latter remains essential for VM/JIT vulnerabilities.
  }%
}
\end{center}

\subsection{RQ4: Ablation Study}
We evaluate how each \textsc{OBsmith} component contributes to bug discovery by enabling exactly one sketch-generation technique or exactly one oracle at a time: the initial LLM sketcher (\textbf{-l}), the feedback loop that refines sketches (\textbf{-f}), automatic sketch extraction from real code (\textbf{-e}), metamorphic testing (\textbf{-m}), and reference-oriented equivalence testing (\textbf{-r}). Every variant is run under the same program budget as the full system. For the feedback setting, we report only \emph{new} bugs found by the refined sketches—if the input already exhibits “bug~$i$” and the loop merely reconfirms bug~$i$, we do not count it.

\paragraph{Results.}
Table~\ref{tab:compare} reports bug classes exposed by each technique. Overall, every component except metamorphic testing contributes unique coverage.

\begin{itemize}
\item \textbf{Reference-oriented equivalence testing (-r)} is the single most impactful oracle. It uncovers the majority of classes tied to \emph{semantic divergence} between the original and obfuscated program—e.g., \emph{eval} crashes, incorrect \emph{undefined} handling, \emph{type errors}, and general \emph{error-handling} inconsistencies. These require cross-version comparison and are rarely surfaced by engine fuzzers.
\item \textbf{LLM-driven generation (-l)} and the \textbf{feedback loop (-f)} each expose multiple \emph{API/semantics–sensitive} issues such as corrupted \emph{constructor.name}, \emph{scope violations}, \emph{silent fixes} (exceptions suppressed or behavior altered without a visible crash), \emph{control-flow changes}, and several \emph{class/async} crashes. The loop frequently converts borderline-invalid or underspecified sketches into valid, higher-tension programs that better stress obfuscators.
\item \textbf{Automatic extraction (-e)} complements learned sketches with patterns mined from real code. Extracted sketches hit bug classes that benefit from idiomatic structures (e.g., class hierarchies and nested control flow), overlapping with but not subsumed by (-l) and (-f).
\item \textbf{Metamorphic testing (-m)} did not reveal additional bug classes under our three generic compiler MRs. This suggests that \emph{obfuscator testing needs domain-specific MRs} (e.g., invariants about identifier renaming, string-table transformations, or control-flow virtualization) rather than the generic arithmetic/logical equivalences commonly used for compilers.
\end{itemize}

\begin{center}
\setlength{\fboxrule}{1pt} 
\setlength{\fboxsep}{6pt} 
\fcolorbox{black!80}{gray!10}{
  \parbox{\dimexpr\columnwidth-2\fboxsep-2\fboxrule\relax}{%
    Answer to RQ4: 
    
    (1) Every component except -m contributes to at least one bug class, with \textbf{-r} providing the largest unique lift. 
    
    (2) \textbf{-l}, \textbf{-f}, and \textbf{-e} provide complementary coverage: LLM-only sketches (-l) already stress obfuscators; feedback (-f) recovers additional crash types (notably \emph{class}); and extraction (-e) adds realistic patterns that trigger control/data-flow issues and reveal async crash.

(3) The negative result for \textbf{-m} suggests that \emph{general} compiler MRs are ill-suited to obfuscator correctness; domain-specific MRs are likely required to make metamorphic testing effective. 
  }%
}
\end{center}




\section{Discussion}
\label{sec:discussion}
Based on investigation from Google~\cite{jiang2025cascade} and WeChat~\cite{li2025jsprotect}, runtime performance is also important in JavaScript obfuscation, so beyond correctness of obfuscators, we also evaluate the obfuscation effect on storage runtime performance. To evaluate OBsmith’s ability to surface these trade-offs, we measured average file size, runtime execution time, and memory usage across multiple obfuscation configurations. Table \ref{tab:obfuscation-performance} summarizes these results.

\begin{table}[htp]
\centering
\caption{Performance comparison of obfuscation configurations with absolute values and percentage changes from original program (filled and enhanced).}
\begin{tabular}{lcccccc}
\toprule
\multirow{2}{*}{Configurations} 
    & \multicolumn{2}{c}{Avg file size (KB)} 
    & \multicolumn{2}{c}{Run time (ms)} 
    & \multicolumn{2}{c}{Memory usage (KB)} \\
\cmidrule(lr){2-3} \cmidrule(lr){4-5} \cmidrule(lr){6-7}
& Abs. & \%$\Delta$ & Abs. & \%$\Delta$ & Abs. & \%$\Delta$ \\
\midrule
Original program & 2.092 & 0.00\% & 28.8 & 0.00\% & 32,311.39 & 0.00\% \\
Obfuscator.IO (Default) & 3.686 & +76.18\% & 28.8 & +0.06\% & 32,385.93 & +0.23\% \\
Obfuscator.IO (Low) & 4.267 & +103.96\% & 29.0 & +0.46\% & 32,464.06 & +0.47\% \\
Obfuscator.IO (Medium) & 15.486 & +640.17\% & 33.5 & +16.04\% & 38,193.16 & +18.20\% \\
Obfuscator.IO (High) & 37.176 & +1,676.89\% & 44.7 & +55.07\% & 43,835.42 & +35.67\% \\
JS-Confuser (Low) & 16.621 & +694.40\% & 29.5 & +2.42\% & 32,517.74 & +0.64\% \\
JS-Confuser (Medium) & 49.097 & +2,246.69\% & 53.6 & +85.95\% & 34,817.05 & +7.75\% \\
JS-Confuser (High) & 159.207 & +7,509.55\% & 113.9 & +295.07\% & 45,408.50 & +40.53\% \\
\bottomrule
\end{tabular}
\label{tab:obfuscation-performance}
\end{table}

Moderate obfuscation settings revealed by OBsmith show manageable costs. When applying Obfuscator.IO (Default, Low) and JS-Confuser (Low) configurations, OBsmith reported only minor runtime overheads (<3\%) and negligible memory differences (<1\%), despite moderate file size increases (76-694\%). These observations demonstrate that OBsmith can reliably detect small but consistent performance shifts under lighter obfuscation.

Aggressive obfuscation highlights OBsmith’s sensitivity to extreme overheads.
OBsmith also exposed the dramatic costs of higher settings. For instance, Obfuscator.IO High inflated file size by +1,676.89\% and runtime by +55.07\%, while JS-Confuser High ballooned file size by +7,509.55\%, runtime by +295.07\%, and memory by +40.53\%. Even the Medium settings produced noticeable slowdowns (+16\% runtime for Obfuscator.IO, +85.95\% for JS-Confuser). These measurements confirm that OBsmith can capture both moderate and extreme performance penalties.

By systematically reporting file size, runtime, and memory metrics across obfuscation configurations, OBsmith provides actionable evidence that stronger obfuscation often comes at the cost of severe inefficiency—insights essential for both researchers and practitioners.

\section{Threats to validity}
As with any empirical study, OBsmith is subject to internal and external threats. 

\textbf{Internal.} (1) OBsmith relies on LLM-generated sketches, which are inherently non-deterministic. Different runs may produce different sketches and uncover different bugs. To mitigate this threat, we use multiple LLMs and generated 100 sketches in each run and report aggregated results. 

(2) The equivalence-checking mechanism may miss subtle semantic differences (e.g., performance or memory behavior). Compared to prior work, we strengthen equivalence checking by incorporating both control-flow validation and scope-level checksums. In addition, we execute programs multiple times to reduce nondeterministic noise. 

(3) The feedback loop introduces another potential source of bias. Such interaction may reinforce trivial variations instead of producing genuinely novel cases. To control for this, we compared bug-finding effectiveness with and without the feedback loop. Our results show that the loop improves bug discovery.

\textbf{External.} OBsmith execute all programs in Node.js engine, which may not fully represent the behavior of JavaScript engines embedded in browsers or alternative run times. Bug manifestation could vary across environments. In addition, although we evaluated multiple widely used open-source obfuscators, our findings may not generalize to all available or proprietary tools. Nevertheless, the chosen obfuscators are representative of current practice in real-world production (based on Google's investigation), and we believe the results provide a meaningful characterization of the reliability of existing obfuscators.

\section{Related Work}
\subsection{JavaScript Obfuscator}
Software obfuscation deliberately transforms code to hinder readability, analysis, and reverse engineering. Common obfuscation techniques include restructuring code, replacing descriptive variable names with unintuitive identifiers, injecting redundant or misleading instructions, manipulating control flow, and encrypting literals such as strings or configuration data~\cite{6185286,zhang2021android}. Although obfuscation can legitimately safeguard proprietary algorithms and sensitive resources (e.g., IP addresses)~\cite{doyle2018privacy,lynn2004positive}, it is also exploited by attackers to mask malicious logic—especially in web and mobile scripts~\cite{brezinski2023metamorphic,281380,10.1016/j.cose.2015.02.007,petrack}. 

Obfuscated code severely diminishes the effectiveness of analysis tools, creating a major obstacle for software testing and static analysis methods~\cite{10.1145/3597926.3598061,10.1109/ICSE-C.2017.79,10.1145/3293882.3330563}. Moreover, obfuscation hampers malware detection by concealing malicious behavior from static analyzers, security filters, machine-learning detectors, and manual reviewers~\cite{pantelaios2024fv8,ren2023jsrevealer,Li2018JSgraphER}. JavaScript, the predominant language for client-side web applications, is especially vulnerable to obfuscation due to its inherently open and accessible nature~\cite{fraunholz2018defending, brewer2010link, 10.5555/2028067.2028070}. 



Prior work on software obfuscation spans obfuscation techniques, detection and analysis methods, and empirical evaluation. Sun et al.~\cite{mate} showed that obfuscation can substantially degrade the effectiveness of anti-malware tooling. Hammad et al.~\cite{10.1145/3180155.3180228} performed a broad experimental evaluation of mainstream anti-malware products under a diverse set of obfuscation strategies, covering multiple open-source, academic, and commercial obfuscators. Complementary to these efforts, researchers have also proposed techniques tailored to detecting obfuscation in JavaScript.

However, for JavaScript obfuscators in particular, the literature offers comparatively limited support for evaluating two developer-facing properties: (1) semantic preservation—whether obfuscation changes program behavior—and (2) performance impact—how transformations affect runtime cost. Skolka et al.~\cite{skolka2019anything} presented an early measurement study of minified and obfuscated web code, but the JavaScript/TypeScript ecosystem and the set of widely deployed obfuscation tools have changed considerably since then. Google’s recent investigation~\cite{jiang2025cascade} identifies Obfuscator.IO as a dominant choice and js-confuser is the second choice, suggesting that conclusions drawn from older tool coverage or test cases may not generalize. These observations motivate revisiting evaluation methodology for modern JavaScript obfuscation and developing automated tests that check both behavioral equivalence and performance regressions.

\subsection{Testing and Sketching.}
Differential testing~\cite{mckeeman1998differential,csmith,wang2023mlirsmith,lu2024understanding,lejit,zang2024jog,yu2025ratte,sharma2023rustsmith} is a software testing technique in which multiple implementations of the same specification are tested using the same set of inputs, with discrepancies in their outputs serving as indicators of potential defects. Csmith~\cite{csmith} is a well-known tool for testing C compilers by randomly generating C programs. It found bugs in main-stream compilers and led to significant attention for compiler testing. Recently, injecting domain knowledge and real-world code has been shown to be effective in compiler testing. Skeletal program enumeration (SPE)~\cite{Skeletal2017} technique to test C/C++ compilers using syntactic skeletons derived from their own regression test-suites and find 217 confirmed bugs in GCC/Clang compiler. Jattack~\cite{jattack} was primarily developed to complement manually-written tests. Developers can embed their knowledge into program generation by specifying holes for exploration, enabling better testing of JIT compilers that require complex structures and execution to reveal bugs.

Recent work has explored complementary approaches to improving software reliability, including static analysis, infrastructure for querying heap objects, and compilation-time optimizations \cite{DingETAL26TypeJinja,ThimmaiahETAL25eStore,AlAwarETAL25Yalla,10795040}. Other efforts develop fuzzers that generate test programs directly \cite{3rdpartyfuzzing,fuzzjit,xia2024fuzz4all,hybridfuzzing}. In contrast, OBsmith uses sketching to encode domain knowledge. Program
sketching~\cite{Solar-LezamaETALCombSketchFinite2006,Solar-LezamaETALStencils2007,SolarLazemaPhD2008,jsketch,psketch,storyboardDS,AlurETAL2013,Singh2015,hong2025effectiveness},
pioneered by the Sketch system~\cite{SolarLazemaPhD2008}, offered an
exciting new advance in scaling program synthesis, where a partial
implementation is given and the goal is to complete
it~\cite{BodikJobstmannAlgoSynthesis2013,CodeHint,Feser2015,Kuncak:2010:CFS:1809028.1806632,Osera:2015:TPS:2813885.2738007,GveroETAL2011,MandelinETAL2005,FengETAL2017,synthesisRecursive,Singh2015}.  EdSketch~\cite{HuaKhurshid2017} and
EdSynth~\cite{YangETAL2017} defined an optimized backtracking search
for completing Java sketches where test executions guided search
pruning.  
The Sketch system provided Java-like Sketch language for writing
partial programs, and deployed SAT and inductive synthesis in a
counterexample-guided loop to complete them.  JSketch enabled
sketching Java programs~\cite{jsketch} by translating Java to Sketch.
PSketch focused on concurrent data structures and enabled sketching
them~\cite{psketch}.

\subsection{Large Language Models}
LLMs have recently demonstrated strong capabilities in diverse code-related tasks, and they have enabled the automation of many software
development and verification themes, including writing
code~\cite{BairiETALFSE2024,XiaETALFSE2025,zhong2025approach,zhong2025april}, clarifying
requirements~\cite{MuETALFSE2024}, software
maintenance~\cite{DilharaETALFSE2024,JinETALFSE2024,JiangETALFSE2024,MaETALICSE2024,XuETALICSE2024,NamETALICSE2024},
software
testing~\cite{RyanETALFSE2024,LiuETALICSE2024,toga,jiang2024generating,legofuzz},
debugging~\cite{WadhwaETALFSE2024,KangETAL2024}, constructing proofs
of theorems in automated provers~\cite{FirstETALFSE2024}, and
human-centric studies~\cite{WangETAL2024,ImranETALICSE2024}. 

Trained on extensive corpora of code and textual data, these models develop an implicit understanding of programming language syntax, semantics, and code structures~\cite{liu2024llm,jiang2025cascade}. Consequently, LLMs exhibit notable capabilities in tasks requiring both natural language comprehension and code manipulation~\cite{jiang2024generating}. While models such as StarCoder~\cite{lozhkov2024starcoder} and Gemini demonstrate proficiency in code generation, they remain susceptible to hallucination, wherein they generate plausible but incorrect or nonsensical outputs~\cite{gambardella-etal-2024-language}. Moreover, LLMs exhibit limitations in tasks demanding precise logical and mathematical reasoning~\cite{zhao2024docmath,yuan2023well}. 

Although LLMs have shown to be effective in the above tasks, there is still a lack of research demonstrating their ability to generate sketches. To the best of our knowledge, OBsmith presents the first study of LLMs in sketching problems for obfuscator testing.  We show that LLMs are capable of generating high-quality sketches thanks to a large training corpus containing domain knowledge.

\section{Conclusion}
In this paper, we present OBsmith, a framework that enables LLM-powered JavaScript obfuscator testing. Using OBsmith, obfuscator developers can use LLMs to generate representative JavaScript sketches which are suitable for testing obfuscators. OBsmith fills sketches and enhances them to enable reference-oriented equivalence testing. OBsmith also exposes program generator as a script to allow developers to write their sketches in the same language as the obfuscators they are testing (JavaScript), enabling them to leverage their domain knowledge to set up a code structure likely to lead to obfuscation problems. Using 600 sketches generated by 6 LLMs, OBsmith generates 3,000 programs and found 11 bugs in 2 popular JavaScript obfuscators used by malware developers. These bugs cover obfuscator crash, obfuscated code crash, control flow changes, etc. 

\section{Data-Availability Statement}
The sketch filling algorithm, enhance algorithm, testing scripts, prompts, sketches, and generated programs for OBsmith are publicly available at \url{https://github.com/shanjiang98/obsmith}.




\begin{acks}
We would like to acknowledge Department of Energy award DE-SC0024467 for their support.
\end{acks}

\appendix

\section{Appendix}
\begin{figure}[htbp]
	\centering
    \begin{lrbox}{0}\begin{minipage}[t]{\linewidth}
\begin{lstlisting}[style = prompt]
<context>
    You are a software testing engineer. Now your task is to generate some template 
    JavaScript programs, and these programs will be used to test JavaScript obfuscators. In 
    template, you should use holes to represent a holder, these holes can be numberLiteral, 
    or booleanLiteral. And these holes can also be numberReference, or booleanReference. And
    you can use binary expression including arithmetic operation(e.g. +, -, *, /,  ...), 
    relation operation(>, <, ==, ===, >=, <=, ...) and logic operation (AND, OR, NOT, ...) 
    in templates, the operands can be any type of literals, references, or holes. I will use 
    templates you generate to do differential testing. Specifically, I will fill holes to 
    get programs and run the filled programs with JavaScript obfuscators and minifiers. 
    Finally I will compare if the transformed JavaScript program have the same output of 
    original filled program.
</context>

<example>
    <template>
        let s1 = numberLiteral;
        let s2 = numberLiteral;
        let b1 = booleanLiteral;
        let arr = [s1++, s2, numberLiteral, numberLiteral, booleanLiteral]
         
        for (int i = 0; i < arr.length; i++) {
          if (logic(relation(numberReference, numberReference, <=),
                    relation(numberReference, numberReference, <=),
                            &&, ||)){
            arr[i] = arithmetic(numberReference, numberReference, +, *);
          }
          if (booleanReference) {
            booleanReference = booleanLiteral;
            booleanReference = booleanReference;
          }
        }
    </template>
    <filled program>
        let s1 = 1098733;
        let s2 = 387923l2;
        let b1 = true;
        let arr = [s1++, s2, 37289, 776, true]
         
        for (int i = 0; i < arr.length; i++) {
          if (arr[1] <= s2 || s2 <= arr[0]){
            arr[i] = arr[0] + s1;
          }
          if (arr[4]){
            b1 = false;
            b1 = arr[4];
          }
        }
    </filled program>
</examples>

<format>
    You should output sketches in a markdown block without any explaination. e.g.:
    ```javascript
        <sketch 1>
        ...
        <sketch 10>
    ```
</format>

<instruction>
    Now please generate <number> different templates for me. You should use templates to 
    cover as many representative javascript programs as possible, including different 
    control flows, common functions, etc. Make sure your output follows format instruction.
</instruction>

\end{lstlisting}
            \caption{OBsmith prompt for LLM-based sketch generation}
    \label{fig:prompt}
 \end{minipage}\end{lrbox}

  \scalebox{0.98}{\usebox{0}}

\end{figure}

\bibliographystyle{ACM-Reference-Format}
\bibliography{main}


\begin{thebibliography}{99}


\ifx \showCODEN    \undefined \def \showCODEN     #1{\unskip}     \fi
\ifx \showISBNx    \undefined \def \showISBNx     #1{\unskip}     \fi
\ifx \showISBNxiii \undefined \def \showISBNxiii  #1{\unskip}     \fi
\ifx \showISSN     \undefined \def \showISSN      #1{\unskip}     \fi
\ifx \showLCCN     \undefined \def \showLCCN      #1{\unskip}     \fi
\ifx \shownote     \undefined \def \shownote      #1{#1}          \fi
\ifx \showarticletitle \undefined \def \showarticletitle #1{#1}   \fi
\ifx \showURL      \undefined \def \showURL       {\relax}        \fi
\providecommand\bibfield[2]{#2}
\providecommand\bibinfo[2]{#2}
\providecommand\natexlab[1]{#1}
\providecommand\showeprint[2][]{arXiv:#2}

\bibitem[bab(2014)]%
        {babel}
 \bibinfo{year}{2014}\natexlab{}.
\newblock \bibinfo{title}{Babel: A tool that helps you write code in the latest version of JavaScript.}
\newblock \bibinfo{howpublished}{\url{https://github.com/babel/babel}}.
\newblock


\bibitem[jsf(2020)]%
        {jsfunfuzz}
 \bibinfo{year}{2020}\natexlab{}.
\newblock \bibinfo{title}{jsfunfuzz}.
\newblock \bibinfo{howpublished}{\url{https://github.com/MozillaSecurity/funfuzz}}.
\newblock


\bibitem[jsc(2022)]%
        {jsconfuser}
 \bibinfo{year}{2022}\natexlab{}.
\newblock \bibinfo{title}{js-confuser}.
\newblock \bibinfo{howpublished}{\url{https://github.com/MichaelXF/JS-Confuser}}.
\newblock


\bibitem[obi(2024)]%
        {obio}
 \bibinfo{year}{2024}\natexlab{}.
\newblock \bibinfo{title}{JavaScript-obfuscator: A Powerful Obfuscator for JavaScript and Node.js}.
\newblock \bibinfo{howpublished}{\url{https://github.com/javascript-obfuscator/javascript-obfuscator}}.
\newblock


\bibitem[Al~Awar et~al\mbox{.}(2025)]%
        {AlAwarETAL25Yalla}
\bibfield{author}{\bibinfo{person}{Nader Al~Awar}, \bibinfo{person}{Zijian Yi}, \bibinfo{person}{George Biros}, {and} \bibinfo{person}{Milos Gligoric}.} \bibinfo{year}{2025}\natexlab{}.
\newblock \showarticletitle{Speeding up the Local {C++} Development Cycle with Header Substitution}. In \bibinfo{booktitle}{\emph{International Symposium on Code Generation and Optimization}}. \bibinfo{pages}{704--717}.
\newblock
\href{https://doi.org/10.1145/3696443.3708942}{doi:\nolinkurl{10.1145/3696443.3708942}}


\bibitem[Alur et~al\mbox{.}(2013)]%
        {AlurETAL2013}
\bibfield{author}{\bibinfo{person}{Rajeev Alur}, \bibinfo{person}{Rastislav Bod{\'{\i}}k}, \bibinfo{person}{Garvit Juniwal}, \bibinfo{person}{Milo M.~K. Martin}, \bibinfo{person}{Mukund Raghothaman}, \bibinfo{person}{Sanjit~A. Seshia}, \bibinfo{person}{Rishabh Singh}, \bibinfo{person}{Armando Solar{-}Lezama}, \bibinfo{person}{Emina Torlak}, {and} \bibinfo{person}{Abhishek Udupa}.} \bibinfo{year}{2013}\natexlab{}.
\newblock \showarticletitle{Syntax-guided synthesis}. In \bibinfo{booktitle}{\emph{FMCAD}}.
\newblock
\href{https://doi.org/10.1109/FMCAD.2013.6679385}{doi:\nolinkurl{10.1109/FMCAD.2013.6679385}}


\bibitem[Bairi et~al\mbox{.}(2024)]%
        {BairiETALFSE2024}
\bibfield{author}{\bibinfo{person}{Ramakrishna Bairi}, \bibinfo{person}{Atharv Sonwane}, \bibinfo{person}{Aditya Kanade}, \bibinfo{person}{Vageesh~D. C.}, \bibinfo{person}{Arun Iyer}, \bibinfo{person}{Suresh Parthasarathy}, \bibinfo{person}{Sriram Rajamani}, \bibinfo{person}{B. Ashok}, {and} \bibinfo{person}{Shashank Shet}.} \bibinfo{year}{2024}\natexlab{}.
\newblock \showarticletitle{CodePlan: Repository-Level Coding using {LLMs} and Planning}.
\newblock  \bibinfo{number}{FSE} (\bibinfo{year}{2024}).
\newblock
\href{https://doi.org/10.1145/3643757}{doi:\nolinkurl{10.1145/3643757}}


\bibitem[Baset et~al\mbox{.}(2017)]%
        {10.1109/ICSE-C.2017.79}
\bibfield{author}{\bibinfo{person}{Salman~A. Baset}, \bibinfo{person}{Shih-Wei Li}, \bibinfo{person}{Philippe Suter}, {and} \bibinfo{person}{Omer Tripp}.} \bibinfo{year}{2017}\natexlab{}.
\newblock \showarticletitle{Identifying Android library dependencies in the presence of code obfuscation and minimization} \emph{(\bibinfo{series}{ICSE-C '17})}.
\newblock
\href{https://doi.org/10.1109/ICSE-C.2017.79}{doi:\nolinkurl{10.1109/ICSE-C.2017.79}}


\bibitem[Blanc et~al\mbox{.}(2012)]%
        {6185286}
\bibfield{author}{\bibinfo{person}{Gregory Blanc}, \bibinfo{person}{Daisuke Miyamoto}, \bibinfo{person}{Mitsuaki Akiyama}, {and} \bibinfo{person}{Youki Kadobayashi}.} \bibinfo{year}{2012}\natexlab{}.
\newblock \showarticletitle{Characterizing Obfuscated JavaScript Using Abstract Syntax Trees: Experimenting with Malicious Scripts}. In \bibinfo{booktitle}{\emph{WAINA}}.
\newblock
\href{https://doi.org/10.1109/WAINA.2012.140}{doi:\nolinkurl{10.1109/WAINA.2012.140}}


\bibitem[Bod{\'{\i}}k and Jobstmann(2013)]%
        {BodikJobstmannAlgoSynthesis2013}
\bibfield{author}{\bibinfo{person}{Rastislav Bod{\'{\i}}k} {and} \bibinfo{person}{Barbara Jobstmann}.} \bibinfo{year}{2013}\natexlab{}.
\newblock \showarticletitle{Algorithmic program synthesis: Introduction}.
\newblock \bibinfo{journal}{\emph{{STTT}}} (\bibinfo{year}{2013}).
\newblock
\href{https://doi.org/10.1007/s10009-013-0287-9}{doi:\nolinkurl{10.1007/s10009-013-0287-9}}


\bibitem[Brewer et~al\mbox{.}(2010)]%
        {brewer2010link}
\bibfield{author}{\bibinfo{person}{Douglas Brewer}, \bibinfo{person}{Kang Li}, \bibinfo{person}{Laksmish Ramaswamy}, {and} \bibinfo{person}{Calton Pu}.} \bibinfo{year}{2010}\natexlab{}.
\newblock \showarticletitle{A link obfuscation service to detect webbots}. In \bibinfo{booktitle}{\emph{SCC}}.
\newblock
\href{https://doi.org/10.1109/SCC.2010.89}{doi:\nolinkurl{10.1109/SCC.2010.89}}


\bibitem[Brezinski and Ferens(2023)]%
        {brezinski2023metamorphic}
\bibfield{author}{\bibinfo{person}{Kenneth Brezinski} {and} \bibinfo{person}{Ken Ferens}.} \bibinfo{year}{2023}\natexlab{}.
\newblock \showarticletitle{Metamorphic malware and obfuscation: a survey of techniques, variants, and generation kits}.
\newblock \bibinfo{journal}{\emph{Security and Communication Networks}} (\bibinfo{year}{2023}).
\newblock
\href{https://doi.org/10.1155/2023/8227751}{doi:\nolinkurl{10.1155/2023/8227751}}


\bibitem[Canfora et~al\mbox{.}(2015)]%
        {canfora2015obfuscation}
\bibfield{author}{\bibinfo{person}{Gerardo Canfora}, \bibinfo{person}{Andrea Di~Sorbo}, \bibinfo{person}{Francesco Mercaldo}, {and} \bibinfo{person}{Corrado~Aaron Visaggio}.} \bibinfo{year}{2015}\natexlab{}.
\newblock \showarticletitle{Obfuscation techniques against signature-based detection: a case study}. In \bibinfo{booktitle}{\emph{2015 Mobile systems technologies workshop (MST)}}.
\newblock
\href{https://doi.org/10.1109/MST.2015.8}{doi:\nolinkurl{10.1109/MST.2015.8}}


\bibitem[Curtsinger et~al\mbox{.}(2011)]%
        {10.5555/2028067.2028070}
\bibfield{author}{\bibinfo{person}{Charlie Curtsinger}, \bibinfo{person}{Benjamin Livshits}, \bibinfo{person}{Benjamin Zorn}, {and} \bibinfo{person}{Christian Seifert}.} \bibinfo{year}{2011}\natexlab{}.
\newblock \showarticletitle{ZOZZLE: fast and precise in-browser JavaScript malware detection} \emph{(\bibinfo{series}{SEC'11})}.
\newblock
\href{https://doi.org/10.5555/2028067.2028070}{doi:\nolinkurl{10.5555/2028067.2028070}}


\bibitem[Dilhara et~al\mbox{.}(2024)]%
        {DilharaETALFSE2024}
\bibfield{author}{\bibinfo{person}{Malinda Dilhara}, \bibinfo{person}{Abhiram Bellur}, \bibinfo{person}{Timofey Bryksin}, {and} \bibinfo{person}{Danny Dig}.} \bibinfo{year}{2024}\natexlab{}.
\newblock \showarticletitle{Unprecedented Code Change Automation: The Fusion of {LLMs} and Transformation by Example}.
\newblock  \bibinfo{number}{{FSE}} (\bibinfo{year}{2024}).
\newblock
\href{https://doi.org/10.1145/3643755}{doi:\nolinkurl{10.1145/3643755}}


\bibitem[Dinella et~al\mbox{.}(2022)]%
        {toga}
\bibfield{author}{\bibinfo{person}{Elizabeth Dinella}, \bibinfo{person}{Gabriel Ryan}, \bibinfo{person}{Todd Mytkowicz}, {and} \bibinfo{person}{Shuvendu~K. Lahiri}.} \bibinfo{year}{2022}\natexlab{}.
\newblock \showarticletitle{{TOGA}: A neural method for test oracle generation}. In \bibinfo{booktitle}{\emph{ICSE}}.
\newblock
\showISBNx{9781450392211}
\href{https://doi.org/10.1145/3510003.3510141}{doi:\nolinkurl{10.1145/3510003.3510141}}


\bibitem[Ding et~al\mbox{.}(2026)]%
        {DingETAL26TypeJinja}
\bibfield{author}{\bibinfo{person}{Cheng Ding}, \bibinfo{person}{Zhong Xu}, \bibinfo{person}{Michael~Y. Levin}, \bibinfo{person}{Wolfram Schulte}, {and} \bibinfo{person}{Milos Gligoric}.} \bibinfo{year}{2026}\natexlab{}.
\newblock \showarticletitle{{TypeJinja}: Static Type Checking of {J}inja Templates at dbt Labs}. In \bibinfo{booktitle}{\emph{International Conference on Software Engineering, Software Engineering in Practice}}.
\newblock
\href{https://doi.org/10.1145/3786583.3786905}{doi:\nolinkurl{10.1145/3786583.3786905}}


\bibitem[Doyle(2018)]%
        {doyle2018privacy}
\bibfield{author}{\bibinfo{person}{Tony Doyle}.} \bibinfo{year}{2018}\natexlab{}.
\newblock \showarticletitle{Privacy, obfuscation, and propertization}.
\newblock \bibinfo{journal}{\emph{IFLA Journal}} (\bibinfo{year}{2018}).
\newblock
\href{https://doi.org/10.1177/0340035218778054}{doi:\nolinkurl{10.1177/0340035218778054}}


\bibitem[Feng et~al\mbox{.}(2017)]%
        {FengETAL2017}
\bibfield{author}{\bibinfo{person}{Yu Feng}, \bibinfo{person}{Ruben Martins}, \bibinfo{person}{Yuepeng Wang}, \bibinfo{person}{Isil Dillig}, {and} \bibinfo{person}{Thomas~W. Reps}.} \bibinfo{year}{2017}\natexlab{}.
\newblock \showarticletitle{Component-based Synthesis for Complex {APIs}}. In \bibinfo{booktitle}{\emph{POPL}}.
\newblock
\href{https://doi.org/10.1145/3009837.3009851}{doi:\nolinkurl{10.1145/3009837.3009851}}


\bibitem[Feser et~al\mbox{.}(2015)]%
        {Feser2015}
\bibfield{author}{\bibinfo{person}{John~K. Feser}, \bibinfo{person}{Swarat Chaudhuri}, {and} \bibinfo{person}{Isil Dillig}.} \bibinfo{year}{2015}\natexlab{}.
\newblock \showarticletitle{Synthesizing Data Structure Transformations from Input-output Examples}. In \bibinfo{booktitle}{\emph{PLDI}}.
\newblock
\href{https://doi.org/10.1145/2737924.2737977}{doi:\nolinkurl{10.1145/2737924.2737977}}


\bibitem[First et~al\mbox{.}(2023)]%
        {FirstETALFSE2024}
\bibfield{author}{\bibinfo{person}{Emily First}, \bibinfo{person}{Markus~N. Rabe}, \bibinfo{person}{Talia Ringer}, {and} \bibinfo{person}{Yuriy Brun}.} \bibinfo{year}{2023}\natexlab{}.
\newblock \showarticletitle{Baldur: Whole-Proof Generation and Repair with Large Language Models}. In \bibinfo{booktitle}{\emph{{ESEC/FSE}}}.
\newblock
\href{https://doi.org/10.1145/3611643.3616243}{doi:\nolinkurl{10.1145/3611643.3616243}}


\bibitem[Fraunholz and Schotten(2018)]%
        {fraunholz2018defending}
\bibfield{author}{\bibinfo{person}{Daniel Fraunholz} {and} \bibinfo{person}{Hans~D. Schotten}.} \bibinfo{year}{2018}\natexlab{}.
\newblock \showarticletitle{Defending Web Servers with Feints, Distraction and Obfuscation}. In \bibinfo{booktitle}{\emph{ICNC}}.
\newblock
\href{https://doi.org/10.1109/ICCNC.2018.8390365}{doi:\nolinkurl{10.1109/ICCNC.2018.8390365}}


\bibitem[Galenson et~al\mbox{.}(2014)]%
        {CodeHint}
\bibfield{author}{\bibinfo{person}{Joel Galenson}, \bibinfo{person}{Philip Reames}, \bibinfo{person}{Rastislav Bodik}, \bibinfo{person}{Bj\"{o}rn Hartmann}, {and} \bibinfo{person}{Koushik Sen}.} \bibinfo{year}{2014}\natexlab{}.
\newblock \showarticletitle{{CodeHint: D}ynamic and Interactive Synthesis of Code Snippets}. In \bibinfo{booktitle}{\emph{ICSE}}.
\newblock
\href{https://doi.org/10.1145/2568225.2568250}{doi:\nolinkurl{10.1145/2568225.2568250}}


\bibitem[Gambardella et~al\mbox{.}(2024)]%
        {gambardella-etal-2024-language}
\bibfield{author}{\bibinfo{person}{Andrew Gambardella}, \bibinfo{person}{Yusuke Iwasawa}, {and} \bibinfo{person}{Yutaka Matsuo}.} \bibinfo{year}{2024}\natexlab{}.
\newblock \showarticletitle{Language Models Do Hard Arithmetic Tasks Easily and Hardly Do Easy Arithmetic Tasks}. In \bibinfo{booktitle}{\emph{ACL}}.
\newblock
\href{https://doi.org/10.18653/v1/2024.acl-short.8}{doi:\nolinkurl{10.18653/v1/2024.acl-short.8}}


\bibitem[Gro{\ss}(2018)]%
        {gross2018fuzzil}
\bibfield{author}{\bibinfo{person}{Samuel Gro{\ss}}.} \bibinfo{year}{2018}\natexlab{}.
\newblock \showarticletitle{Fuzzil: Coverage guided fuzzing for javascript engines}.
\newblock \bibinfo{journal}{\emph{Department of Informatics, Karlsruhe Institute of Technology}} (\bibinfo{year}{2018}).
\newblock


\bibitem[Gvero et~al\mbox{.}(2011)]%
        {GveroETAL2011}
\bibfield{author}{\bibinfo{person}{Tihomir Gvero}, \bibinfo{person}{Viktor Kuncak}, {and} \bibinfo{person}{Ruzica Piskac}.} \bibinfo{year}{2011}\natexlab{}.
\newblock \showarticletitle{Interactive Synthesis of Code Snippets}. In \bibinfo{booktitle}{\emph{CAV}}.
\newblock
\href{https://doi.org/10.5555/2032305.2032338}{doi:\nolinkurl{10.5555/2032305.2032338}}


\bibitem[Hammad et~al\mbox{.}(2018)]%
        {10.1145/3180155.3180228}
\bibfield{author}{\bibinfo{person}{Mahmoud Hammad}, \bibinfo{person}{Joshua Garcia}, {and} \bibinfo{person}{Sam Malek}.} \bibinfo{year}{2018}\natexlab{}.
\newblock \showarticletitle{A large-scale empirical study on the effects of code obfuscations on Android apps and anti-malware products} \emph{(\bibinfo{series}{ICSE})}.
\newblock
\href{https://doi.org/10.1145/3180155.3180228}{doi:\nolinkurl{10.1145/3180155.3180228}}


\bibitem[Hong et~al\mbox{.}(2025)]%
        {hong2025effectiveness}
\bibfield{author}{\bibinfo{person}{Yang Hong}, \bibinfo{person}{Shan Jiang}, \bibinfo{person}{Yulei Fu}, {and} \bibinfo{person}{Sarfraz Khurshid}.} \bibinfo{year}{2025}\natexlab{}.
\newblock \showarticletitle{On the Effectiveness of Large Language Models in Writing Alloy Formulas}.
\newblock \bibinfo{journal}{\emph{arXiv preprint arXiv:2502.15441}} (\bibinfo{year}{2025}).
\newblock
\href{https://doi.org/10.48550/arXiv.2502.15441}{doi:\nolinkurl{10.48550/arXiv.2502.15441}}


\bibitem[Hua and Khurshid(2017)]%
        {HuaKhurshid2017}
\bibfield{author}{\bibinfo{person}{Jinru Hua} {and} \bibinfo{person}{Sarfraz Khurshid}.} \bibinfo{year}{2017}\natexlab{}.
\newblock \showarticletitle{{EdSketch}: Execution-Driven Sketching for {Java}}. In \bibinfo{booktitle}{\emph{{SPIN}}}.
\newblock
\href{https://doi.org/10.1145/3092282.3092285}{doi:\nolinkurl{10.1145/3092282.3092285}}


\bibitem[Imran et~al\mbox{.}(2024)]%
        {ImranETALICSE2024}
\bibfield{author}{\bibinfo{person}{Mia~Mohammad Imran}, \bibinfo{person}{Preetha Chatterjee}, {and} \bibinfo{person}{Kostadin Damevski}.} \bibinfo{year}{2024}\natexlab{}.
\newblock \showarticletitle{Uncovering the Causes of Emotions in Software Developer Communication Using Zero-shot {LLMs}}. In \bibinfo{booktitle}{\emph{ICSE}}.
\newblock
\href{https://doi.org/10.1145/3597503.3639223}{doi:\nolinkurl{10.1145/3597503.3639223}}


\bibitem[Jeon et~al\mbox{.}(2015)]%
        {jsketch}
\bibfield{author}{\bibinfo{person}{Jinseong Jeon}, \bibinfo{person}{Xiaokang Qiu}, \bibinfo{person}{Jeffrey~S. Foster}, {and} \bibinfo{person}{Armando Solar-Lezama}.} \bibinfo{year}{2015}\natexlab{}.
\newblock \showarticletitle{{JSketch}: Sketching for {Java}}. In \bibinfo{booktitle}{\emph{FSE}}.
\newblock
\href{https://doi.org/10.1145/2786805.2803189}{doi:\nolinkurl{10.1145/2786805.2803189}}


\bibitem[Jiang et~al\mbox{.}(2023a)]%
        {3rdpartyfuzzing}
\bibfield{author}{\bibinfo{person}{Ling Jiang}, \bibinfo{person}{Hengchen Yuan}, \bibinfo{person}{Qiyi Tang}, \bibinfo{person}{Sen Nie}, \bibinfo{person}{Shi Wu}, {and} \bibinfo{person}{Yuqun Zhang}.} \bibinfo{year}{2023}\natexlab{a}.
\newblock \showarticletitle{Third-Party Library Dependency for Large-Scale SCA in the C/C++ Ecosystem: How Far Are We?}. In \bibinfo{booktitle}{\emph{ISSTA}}.
\newblock
\href{https://doi.org/10.1145/3597926.3598143}{doi:\nolinkurl{10.1145/3597926.3598143}}


\bibitem[Jiang et~al\mbox{.}(2023b)]%
        {hybridfuzzing}
\bibfield{author}{\bibinfo{person}{Ling Jiang}, \bibinfo{person}{Hengchen Yuan}, \bibinfo{person}{Mingyuan Wu}, \bibinfo{person}{Lingming Zhang}, {and} \bibinfo{person}{Yuqun Zhang}.} \bibinfo{year}{2023}\natexlab{b}.
\newblock \showarticletitle{Evaluating and Improving Hybrid Fuzzing}. In \bibinfo{booktitle}{\emph{ICSE}}.
\newblock
\href{https://doi.org/10.1109/ICSE48619.2023.00045}{doi:\nolinkurl{10.1109/ICSE48619.2023.00045}}


\bibitem[Jiang et~al\mbox{.}(2026)]%
        {jiang2025cascade}
\bibfield{author}{\bibinfo{person}{Shan Jiang}, \bibinfo{person}{Pranoy Kovuri}, \bibinfo{person}{David Tao}, {and} \bibinfo{person}{Zhixun Tan}.} \bibinfo{year}{2026}\natexlab{}.
\newblock \showarticletitle{CASCADE: LLM-Powered JavaScript Deobfuscator at Google}. In \bibinfo{booktitle}{\emph{International Conference on Software Engineering, Software Engineering in Practice}}.
\newblock
\href{https://doi.org/10.1145/3786583.3786873}{doi:\nolinkurl{10.1145/3786583.3786873}}


\bibitem[Jiang et~al\mbox{.}(2024b)]%
        {jiang2024generating}
\bibfield{author}{\bibinfo{person}{Shan Jiang}, \bibinfo{person}{Chenguang Zhu}, {and} \bibinfo{person}{Sarfraz Khurshid}.} \bibinfo{year}{2024}\natexlab{b}.
\newblock \showarticletitle{Generating executable oracles to check conformance of client code to requirements of JDK Javadocs using LLMs}.
\newblock \bibinfo{journal}{\emph{arXiv preprint arXiv:2411.01789}} (\bibinfo{year}{2024}).
\newblock
\href{https://doi.org/10.48550/arXiv.2411.01789}{doi:\nolinkurl{10.48550/arXiv.2411.01789}}


\bibitem[Jiang et~al\mbox{.}(2024a)]%
        {JiangETALFSE2024}
\bibfield{author}{\bibinfo{person}{Zhihan Jiang}, \bibinfo{person}{Jinyang Liu}, \bibinfo{person}{Zhuangbin Chen}, \bibinfo{person}{Yichen Li}, \bibinfo{person}{Junjie Huang}, \bibinfo{person}{Yintong Huo}, \bibinfo{person}{Pinjia He}, \bibinfo{person}{Jiazhen Gu}, {and} \bibinfo{person}{Michael~R. Lyu}.} \bibinfo{year}{2024}\natexlab{a}.
\newblock \showarticletitle{{LILAC}: Log Parsing using {LLMs} with Adaptive Parsing Cache}.
\newblock  \bibinfo{number}{FSE} (\bibinfo{year}{2024}).
\newblock
\href{https://doi.org/10.1145/3643733}{doi:\nolinkurl{10.1145/3643733}}


\bibitem[Jin and Lin(2024)]%
        {JinETALFSE2024}
\bibfield{author}{\bibinfo{person}{Xin Jin} {and} \bibinfo{person}{Zhiqiang Lin}.} \bibinfo{year}{2024}\natexlab{}.
\newblock \showarticletitle{{SimLLM}: Calculating Semantic Similarity in Code Summaries using a Large Language Model-Based Approach}.
\newblock  \bibinfo{number}{FSE} (\bibinfo{year}{2024}).
\newblock
\href{https://doi.org/10.1145/3660769}{doi:\nolinkurl{10.1145/3660769}}


\bibitem[Kang et~al\mbox{.}(2024)]%
        {KangETAL2024}
\bibfield{author}{\bibinfo{person}{Sungmin Kang}, \bibinfo{person}{Gabin An}, {and} \bibinfo{person}{Shin Yoo}.} \bibinfo{year}{2024}\natexlab{}.
\newblock \showarticletitle{A Quantitative and Qualitative Evaluation of {LLM}-Based Explainable Fault Localization}.
\newblock  \bibinfo{number}{FSE} (\bibinfo{year}{2024}).
\newblock
\href{https://doi.org/10.1145/3660771}{doi:\nolinkurl{10.1145/3660771}}


\bibitem[Kneuss et~al\mbox{.}(2013)]%
        {synthesisRecursive}
\bibfield{author}{\bibinfo{person}{Etienne Kneuss}, \bibinfo{person}{Ivan Kuraj}, \bibinfo{person}{Viktor Kuncak}, {and} \bibinfo{person}{Philippe Suter}.} \bibinfo{year}{2013}\natexlab{}.
\newblock \showarticletitle{Synthesis Modulo Recursive Functions}. In \bibinfo{booktitle}{\emph{OOPSLA}}.
\newblock
\href{https://doi.org/10.1145/2544173.2509555}{doi:\nolinkurl{10.1145/2544173.2509555}}


\bibitem[Kuncak et~al\mbox{.}(2010)]%
        {Kuncak:2010:CFS:1809028.1806632}
\bibfield{author}{\bibinfo{person}{Viktor Kuncak}, \bibinfo{person}{Mika\"{e}l Mayer}, \bibinfo{person}{Ruzica Piskac}, {and} \bibinfo{person}{Philippe Suter}.} \bibinfo{year}{2010}\natexlab{}.
\newblock \showarticletitle{Complete Functional Synthesis}. In \bibinfo{booktitle}{\emph{PLDI}}.
\newblock
\href{https://doi.org/10.1145/1809028.1806632}{doi:\nolinkurl{10.1145/1809028.1806632}}


\bibitem[Li et~al\mbox{.}(2018)]%
        {Li2018JSgraphER}
\bibfield{author}{\bibinfo{person}{Bo Li}, \bibinfo{person}{Phani Vadrevu}, \bibinfo{person}{Kyu~Hyung Lee}, {and} \bibinfo{person}{Roberto Perdisci}.} \bibinfo{year}{2018}\natexlab{}.
\newblock \showarticletitle{JSgraph: Enabling Reconstruction of Web Attacks via Efficient Tracking of Live In-Browser JavaScript Executions}. In \bibinfo{booktitle}{\emph{NDSS}}.
\newblock
\href{https://doi.org/10.14722/ndss.2018.23319}{doi:\nolinkurl{10.14722/ndss.2018.23319}}


\bibitem[Li et~al\mbox{.}(2021)]%
        {unifuzz-li}
\bibfield{author}{\bibinfo{person}{Yuwei Li}, \bibinfo{person}{Shouling Ji}, \bibinfo{person}{Yuan Chen}, \bibinfo{person}{Sizhuang Liang}, \bibinfo{person}{Wei-Han Lee}, \bibinfo{person}{Yueyao Chen}, \bibinfo{person}{Chenyang Lyu}, \bibinfo{person}{Chunming Wu}, \bibinfo{person}{Raheem Beyah}, \bibinfo{person}{Peng Cheng}, \bibinfo{person}{Kangjie Lu}, {and} \bibinfo{person}{Ting Wang}.} \bibinfo{year}{2021}\natexlab{}.
\newblock \showarticletitle{{UNIFUZZ}: A Holistic and Pragmatic Metrics-Driven Platform for Evaluating Fuzzers}. In \bibinfo{booktitle}{\emph{{USENIX} Security}} \emph{(\bibinfo{series}{SEC '21})}.
\newblock
\urldef\tempurl%
\url{https://www.usenix.org/conference/usenixsecurity21/presentation/li-yuwei}
\showURL{%
\tempurl}


\bibitem[Li et~al\mbox{.}(2025)]%
        {li2025jsprotect}
\bibfield{author}{\bibinfo{person}{Zhihao Li}, \bibinfo{person}{Chaozheng Wang}, \bibinfo{person}{Zongjie Li}, \bibinfo{person}{Xinyong Peng}, \bibinfo{person}{Zelin Su}, \bibinfo{person}{Qun Xia}, \bibinfo{person}{Haochuan Lu}, \bibinfo{person}{Ting Xiong}, \bibinfo{person}{Man~Ho Lam}, \bibinfo{person}{Shuzheng Gao}, {et~al\mbox{.}}} \bibinfo{year}{2025}\natexlab{}.
\newblock \showarticletitle{JSProtect: A Scalable Obfuscation Framework for Mini-Games in WeChat}.
\newblock \bibinfo{journal}{\emph{arXiv preprint arXiv:2509.24498}} (\bibinfo{year}{2025}).
\newblock
\href{https://doi.org/10.48550/arXiv.2509.24498}{doi:\nolinkurl{10.48550/arXiv.2509.24498}}


\bibitem[Liu et~al\mbox{.}(2025)]%
        {liu2024llm}
\bibfield{author}{\bibinfo{person}{Kaibo Liu}, \bibinfo{person}{Zhenpeng Chen}, \bibinfo{person}{Yiyang Liu}, \bibinfo{person}{Jie~M. Zhang}, \bibinfo{person}{Mark Harman}, \bibinfo{person}{Yudong Han}, \bibinfo{person}{Yun Ma}, \bibinfo{person}{Yihong Dong}, \bibinfo{person}{Ge Li}, {and} \bibinfo{person}{Gang Huang}.} \bibinfo{year}{2025}\natexlab{}.
\newblock \showarticletitle{{LLM}-Powered Test Case Generation for Detecting Bugs in Plausible Programs}. In \bibinfo{booktitle}{\emph{ACL}}.
\newblock
\href{https://doi.org/10.18653/v1/2025.acl-long.20}{doi:\nolinkurl{10.18653/v1/2025.acl-long.20}}


\bibitem[Liu et~al\mbox{.}(2024)]%
        {LiuETALICSE2024}
\bibfield{author}{\bibinfo{person}{Zhe Liu}, \bibinfo{person}{Chunyang Chen}, \bibinfo{person}{Junjie Wang}, \bibinfo{person}{Mengzhuo Chen}, \bibinfo{person}{Boyu Wu}, \bibinfo{person}{Xing Che}, \bibinfo{person}{Dandan Wang}, {and} \bibinfo{person}{Qing Wang}.} \bibinfo{year}{2024}\natexlab{}.
\newblock \showarticletitle{Make {LLM} a Testing Expert: Bringing Human-like Interaction to Mobile GUI Testing via Functionality-aware Decisions}. In \bibinfo{booktitle}{\emph{ICSE}}.
\newblock
\href{https://doi.org/10.1145/3597503.3639180}{doi:\nolinkurl{10.1145/3597503.3639180}}


\bibitem[Lozhkov et~al\mbox{.}(2024)]%
        {lozhkov2024starcoder}
\bibfield{author}{\bibinfo{person}{Anton Lozhkov}, \bibinfo{person}{Raymond Li}, \bibinfo{person}{Loubna~Ben Allal}, \bibinfo{person}{Federico Cassano}, \bibinfo{person}{Joel Lamy-Poirier}, \bibinfo{person}{Nouamane Tazi}, \bibinfo{person}{Ao Tang}, \bibinfo{person}{Dmytro Pykhtar}, \bibinfo{person}{Jiawei Liu}, \bibinfo{person}{Yuxiang Wei}, {et~al\mbox{.}}} \bibinfo{year}{2024}\natexlab{}.
\newblock \showarticletitle{Starcoder 2 and the stack v2: The next generation}.
\newblock \bibinfo{journal}{\emph{arXiv preprint arXiv:2402.19173}} (\bibinfo{year}{2024}).
\newblock
\href{https://doi.org/10.48550/arXiv.2402.19173}{doi:\nolinkurl{10.48550/arXiv.2402.19173}}


\bibitem[Lu et~al\mbox{.}(2024)]%
        {lu2024understanding}
\bibfield{author}{\bibinfo{person}{Yifei Lu}, \bibinfo{person}{Weidong Hou}, \bibinfo{person}{Minxue Pan}, \bibinfo{person}{Xuandong Li}, {and} \bibinfo{person}{Zhendong Su}.} \bibinfo{year}{2024}\natexlab{}.
\newblock \showarticletitle{Understanding and finding Java decompiler bugs}.
\newblock  \bibinfo{number}{OOPSLA} (\bibinfo{year}{2024}).
\newblock
\href{https://doi.org/10.1145/3649860}{doi:\nolinkurl{10.1145/3649860}}


\bibitem[Lynn et~al\mbox{.}(2004)]%
        {lynn2004positive}
\bibfield{author}{\bibinfo{person}{Benjamin Lynn}, \bibinfo{person}{Manoj Prabhakaran}, {and} \bibinfo{person}{Amit Sahai}.} \bibinfo{year}{2004}\natexlab{}.
\newblock \showarticletitle{Positive results and techniques for obfuscation}. In \bibinfo{booktitle}{\emph{EUROCRYPT}}. Springer.
\newblock
\href{https://doi.org/10.1007/978-3-540-24676-3_2}{doi:\nolinkurl{10.1007/978-3-540-24676-3_2}}


\bibitem[Ma et~al\mbox{.}(2024)]%
        {MaETALICSE2024}
\bibfield{author}{\bibinfo{person}{Zeyang Ma}, \bibinfo{person}{An~Ran Chen}, \bibinfo{person}{Dong~Jae Kim}, \bibinfo{person}{Tse-Hsun Chen}, {and} \bibinfo{person}{Shaowei Wang}.} \bibinfo{year}{2024}\natexlab{}.
\newblock \showarticletitle{{LLMParser}: An Exploratory Study on Using Large Language Models for Log Parsing}. In \bibinfo{booktitle}{\emph{ICSE}}.
\newblock
\href{https://doi.org/10.1145/3597503.3639150}{doi:\nolinkurl{10.1145/3597503.3639150}}


\bibitem[Maiorca et~al\mbox{.}(2015)]%
        {10.1016/j.cose.2015.02.007}
\bibfield{author}{\bibinfo{person}{Davide Maiorca}, \bibinfo{person}{Davide Ariu}, \bibinfo{person}{Igino Corona}, \bibinfo{person}{Marco Aresu}, {and} \bibinfo{person}{Giorgio Giacinto}.} \bibinfo{year}{2015}\natexlab{}.
\newblock \showarticletitle{Stealth attacks: An extended insight into the obfuscation effects on Android malware}.
\newblock \bibinfo{journal}{\emph{Computers \& Security}} (\bibinfo{year}{2015}).
\newblock
\href{https://doi.org/10.1016/j.cose.2015.02.007}{doi:\nolinkurl{10.1016/j.cose.2015.02.007}}


\bibitem[Mandelin et~al\mbox{.}(2005)]%
        {MandelinETAL2005}
\bibfield{author}{\bibinfo{person}{David Mandelin}, \bibinfo{person}{Lin Xu}, \bibinfo{person}{Rastislav Bod\'{\i}k}, {and} \bibinfo{person}{Doug Kimelman}.} \bibinfo{year}{2005}\natexlab{}.
\newblock \showarticletitle{Jungloid mining: helping to navigate the API jungle}. In \bibinfo{booktitle}{\emph{PLDI}}. \bibinfo{numpages}{14}~pages.
\newblock
\href{https://doi.org/10.1145/1064978.1065018}{doi:\nolinkurl{10.1145/1064978.1065018}}


\bibitem[McKeeman(1998)]%
        {mckeeman1998differential}
\bibfield{author}{\bibinfo{person}{William~M McKeeman}.} \bibinfo{year}{1998}\natexlab{}.
\newblock \showarticletitle{Differential testing for software}.
\newblock \bibinfo{journal}{\emph{Digital Technical Journal}} (\bibinfo{year}{1998}).
\newblock


\bibitem[Meng et~al\mbox{.}(2022)]%
        {meng2022batmapper}
\bibfield{author}{\bibinfo{person}{Chuize Meng}, \bibinfo{person}{Shan Jiang}, \bibinfo{person}{Mengning Wu}, \bibinfo{person}{Xuan Xiao}, \bibinfo{person}{Dan Tao}, {and} \bibinfo{person}{Ruipeng Gao}.} \bibinfo{year}{2022}\natexlab{}.
\newblock \showarticletitle{BatMapper-Plus: Smartphone-Based Multi-level Indoor Floor Plan Construction via Acoustic Ranging and Inertial Sensing}. In \bibinfo{booktitle}{\emph{WASA}}.
\newblock
\href{https://doi.org/10.1007/978-3-031-19214-2_13}{doi:\nolinkurl{10.1007/978-3-031-19214-2_13}}


\bibitem[Meng et~al\mbox{.}(2023)]%
        {meng2023smartphone}
\bibfield{author}{\bibinfo{person}{Chuize Meng}, \bibinfo{person}{Shan Jiang}, \bibinfo{person}{Mengning Wu}, \bibinfo{person}{Xuan Xiao}, \bibinfo{person}{Dan Tao}, {and} \bibinfo{person}{Ruipeng Gao}.} \bibinfo{year}{2023}\natexlab{}.
\newblock \showarticletitle{Smartphone-Based Indoor Floor Plan Construction via Acoustic Ranging and Inertial Tracking}.
\newblock \bibinfo{journal}{\emph{Machines}} (\bibinfo{year}{2023}).
\newblock
\href{https://doi.org/10.3390/machines11020205}{doi:\nolinkurl{10.3390/machines11020205}}


\bibitem[Mu et~al\mbox{.}(2024)]%
        {MuETALFSE2024}
\bibfield{author}{\bibinfo{person}{Fangwen Mu}, \bibinfo{person}{Lin Shi}, \bibinfo{person}{Song Wang}, \bibinfo{person}{Zhuohao Yu}, \bibinfo{person}{Binquan Zhang}, \bibinfo{person}{ChenXue Wang}, \bibinfo{person}{Shichao Liu}, {and} \bibinfo{person}{Qing Wang}.} \bibinfo{year}{2024}\natexlab{}.
\newblock \showarticletitle{ClarifyGPT: A Framework for Enhancing {LLM}-Based Code Generation via Requirements Clarification}.
\newblock  \bibinfo{number}{FSE} (\bibinfo{year}{2024}).
\newblock
\href{https://doi.org/10.1145/3660810}{doi:\nolinkurl{10.1145/3660810}}


\bibitem[Nam et~al\mbox{.}(2024)]%
        {NamETALICSE2024}
\bibfield{author}{\bibinfo{person}{Daye Nam}, \bibinfo{person}{Andrew Macvean}, \bibinfo{person}{Vincent Hellendoorn}, \bibinfo{person}{Bogdan Vasilescu}, {and} \bibinfo{person}{Brad Myers}.} \bibinfo{year}{2024}\natexlab{}.
\newblock \showarticletitle{Using an {LLM} to Help With Code Understanding}. In \bibinfo{booktitle}{\emph{ICSE}}.
\newblock
\href{https://doi.org/10.1145/3597503.3639187}{doi:\nolinkurl{10.1145/3597503.3639187}}


\bibitem[Ni and Li(2025)]%
        {legofuzz}
\bibfield{author}{\bibinfo{person}{Yunbo Ni} {and} \bibinfo{person}{Shaohua Li}.} \bibinfo{year}{2025}\natexlab{}.
\newblock \showarticletitle{Interleaving Large Language Models for Compiler Testing}.
\newblock \bibinfo{journal}{\emph{Proc. ACM Program. Lang.}} \bibinfo{number}{OOPSLA2} (\bibinfo{year}{2025}).
\newblock
\href{https://doi.org/10.1145/3763079}{doi:\nolinkurl{10.1145/3763079}}


\bibitem[Osera and Zdancewic(2015)]%
        {Osera:2015:TPS:2813885.2738007}
\bibfield{author}{\bibinfo{person}{Peter-Michael Osera} {and} \bibinfo{person}{Steve Zdancewic}.} \bibinfo{year}{2015}\natexlab{}.
\newblock \showarticletitle{Type-and-example-directed Program Synthesis}. In \bibinfo{booktitle}{\emph{PLDI}}.
\newblock
\href{https://doi.org/10.1145/2737924.2738007}{doi:\nolinkurl{10.1145/2737924.2738007}}


\bibitem[Pantelaios and Kapravelos(2024)]%
        {pantelaios2024fv8}
\bibfield{author}{\bibinfo{person}{Nikolaos Pantelaios} {and} \bibinfo{person}{Alexandros Kapravelos}.} \bibinfo{year}{2024}\natexlab{}.
\newblock \showarticletitle{{FV8}: A Forced Execution {JavaScript} Engine for Detecting Evasive Techniques}. In \bibinfo{booktitle}{\emph{USENIX Security 24}}.
\newblock
\urldef\tempurl%
\url{https://www.usenix.org/conference/usenixsecurity24/presentation/pantelaios}
\showURL{%
\tempurl}


\bibitem[Park et~al\mbox{.}(2020)]%
        {park:die}
\bibfield{author}{\bibinfo{person}{Soyeon Park}, \bibinfo{person}{Wen Xu}, \bibinfo{person}{Insu Yun}, \bibinfo{person}{Daehee Jang}, {and} \bibinfo{person}{Taesoo Kim}.} \bibinfo{year}{2020}\natexlab{}.
\newblock \showarticletitle{{Fuzzing JavaScript Engines with Aspect-preserving Mutation}}. In \bibinfo{booktitle}{\emph{IEEE Symposium on Security and Privacy (Oakland)}}.
\newblock
\href{https://doi.org/10.1109/SP40000.2020.00067}{doi:\nolinkurl{10.1109/SP40000.2020.00067}}


\bibitem[Raychev et~al\mbox{.}(2016)]%
        {data150k}
\bibfield{author}{\bibinfo{person}{Veselin Raychev}, \bibinfo{person}{Pavol Bielik}, \bibinfo{person}{Martin Vechev}, {and} \bibinfo{person}{Andreas Krause}.} \bibinfo{year}{2016}\natexlab{}.
\newblock \showarticletitle{Learning programs from noisy data}.
\newblock \bibinfo{journal}{\emph{POPL}} (\bibinfo{year}{2016}).
\newblock
\href{https://doi.org/10.1145/2837614.2837671}{doi:\nolinkurl{10.1145/2837614.2837671}}


\bibitem[Ren et~al\mbox{.}(2023)]%
        {ren2023jsrevealer}
\bibfield{author}{\bibinfo{person}{Kunlun Ren}, \bibinfo{person}{Weizhong Qiang}, \bibinfo{person}{Yueming Wu}, \bibinfo{person}{Yi Zhou}, \bibinfo{person}{Deqing Zou}, {and} \bibinfo{person}{Hai Jin}.} \bibinfo{year}{2023}\natexlab{}.
\newblock \showarticletitle{JSRevealer: A Robust Malicious JavaScript Detector against Obfuscation}. In \bibinfo{booktitle}{\emph{DSN}}.
\newblock
\href{https://doi.org/10.1109/DSN58367.2023.00041}{doi:\nolinkurl{10.1109/DSN58367.2023.00041}}


\bibitem[Ren et~al\mbox{.}(2022)]%
        {petrack}
\bibfield{author}{\bibinfo{person}{Xiaotong Ren}, \bibinfo{person}{Shuli Zhu}, \bibinfo{person}{Chuize Meng}, \bibinfo{person}{Shan Jiang}, \bibinfo{person}{Xuan Xiao}, \bibinfo{person}{Dan Tao}, {and} \bibinfo{person}{Ruipeng Gao}.} \bibinfo{year}{2022}\natexlab{}.
\newblock \showarticletitle{PeTrack: Smartphone-based Pedestrian Tracking in Underground Parking Lot}. In \bibinfo{booktitle}{\emph{MSN}}.
\newblock
\href{https://doi.org/10.1109/MSN57253.2022.00122}{doi:\nolinkurl{10.1109/MSN57253.2022.00122}}


\bibitem[Ryan et~al\mbox{.}(2024)]%
        {RyanETALFSE2024}
\bibfield{author}{\bibinfo{person}{Gabriel Ryan}, \bibinfo{person}{Siddhartha Jain}, \bibinfo{person}{Mingyue Shang}, \bibinfo{person}{Shiqi Wang}, \bibinfo{person}{Xiaofei Ma}, \bibinfo{person}{Murali~Krishna Ramanathan}, {and} \bibinfo{person}{Baishakhi Ray}.} \bibinfo{year}{2024}\natexlab{}.
\newblock \showarticletitle{Code-Aware Prompting: A Study of Coverage-Guided Test Generation in Regression Setting using {LLM}}.
\newblock  \bibinfo{number}{FSE} (\bibinfo{year}{2024}).
\newblock
\href{https://doi.org/10.1145/3643769}{doi:\nolinkurl{10.1145/3643769}}


\bibitem[Schloegel et~al\mbox{.}(2022)]%
        {281380}
\bibfield{author}{\bibinfo{person}{Moritz Schloegel}, \bibinfo{person}{Tim Blazytko}, \bibinfo{person}{Moritz Contag}, \bibinfo{person}{Cornelius Aschermann}, \bibinfo{person}{Julius Basler}, \bibinfo{person}{Thorsten Holz}, {and} \bibinfo{person}{Ali Abbasi}.} \bibinfo{year}{2022}\natexlab{}.
\newblock \showarticletitle{Loki: Hardening Code Obfuscation Against Automated Attacks}. In \bibinfo{booktitle}{\emph{SEC}}.
\newblock
\urldef\tempurl%
\url{https://www.usenix.org/conference/usenixsecurity22/presentation/schloegel}
\showURL{%
\tempurl}


\bibitem[Sharma et~al\mbox{.}(2023)]%
        {sharma2023rustsmith}
\bibfield{author}{\bibinfo{person}{Mayank Sharma}, \bibinfo{person}{Pingshi Yu}, {and} \bibinfo{person}{Alastair~F Donaldson}.} \bibinfo{year}{2023}\natexlab{}.
\newblock \showarticletitle{Rustsmith: Random differential compiler testing for rust}. In \bibinfo{booktitle}{\emph{ISSTA}}.
\newblock
\href{https://doi.org/10.1145/3597926.3604919}{doi:\nolinkurl{10.1145/3597926.3604919}}


\bibitem[Singh and Gulwani(2015)]%
        {Singh2015}
\bibfield{author}{\bibinfo{person}{Rishabh Singh} {and} \bibinfo{person}{Sumit Gulwani}.} \bibinfo{year}{2015}\natexlab{}.
\newblock \showarticletitle{Predicting a Correct Program in Programming by Example}. In \bibinfo{booktitle}{\emph{{CAV}}}.
\newblock
\href{https://doi.org/10.1007/978-3-319-21690-4_23}{doi:\nolinkurl{10.1007/978-3-319-21690-4_23}}


\bibitem[Singh and Solar-Lezama(2011)]%
        {storyboardDS}
\bibfield{author}{\bibinfo{person}{Rishabh Singh} {and} \bibinfo{person}{Armando Solar-Lezama}.} \bibinfo{year}{2011}\natexlab{}.
\newblock \showarticletitle{Synthesizing Data Structure Manipulations from Storyboards}. In \bibinfo{booktitle}{\emph{{FSE}}}.
\newblock
\href{https://doi.org/10.1145/2025113.2025153}{doi:\nolinkurl{10.1145/2025113.2025153}}


\bibitem[Skolka et~al\mbox{.}(2019)]%
        {skolka2019anything}
\bibfield{author}{\bibinfo{person}{Philippe Skolka}, \bibinfo{person}{Cristian-Alexandru Staicu}, {and} \bibinfo{person}{Michael Pradel}.} \bibinfo{year}{2019}\natexlab{}.
\newblock \showarticletitle{Anything to hide? studying minified and obfuscated code in the web}. In \bibinfo{booktitle}{\emph{WWW}}.
\newblock
\href{https://doi.org/10.1145/3308558.3313752}{doi:\nolinkurl{10.1145/3308558.3313752}}


\bibitem[Solar-Lezama(2008)]%
        {SolarLazemaPhD2008}
\bibfield{author}{\bibinfo{person}{Armando Solar-Lezama}.} \bibinfo{year}{2008}\natexlab{}.
\newblock \emph{\bibinfo{title}{Program Synthesis by Sketching}}.
\newblock \bibinfo{thesistype}{Ph.\,D. Dissertation}. \bibinfo{school}{University of California, Berkeley}.
\newblock


\bibitem[Solar-Lezama et~al\mbox{.}(2007)]%
        {Solar-LezamaETALStencils2007}
\bibfield{author}{\bibinfo{person}{Armando Solar-Lezama}, \bibinfo{person}{Gilad Arnold}, \bibinfo{person}{Liviu Tancau}, \bibinfo{person}{Rastislav Bodik}, \bibinfo{person}{Vijay Saraswat}, {and} \bibinfo{person}{Sanjit Seshia}.} \bibinfo{year}{2007}\natexlab{}.
\newblock \showarticletitle{Sketching Stencils}.
\newblock \bibinfo{journal}{\emph{{PLDI}}} (\bibinfo{year}{2007}).
\newblock
\href{https://doi.org/10.1145/1250734.1250754}{doi:\nolinkurl{10.1145/1250734.1250754}}


\bibitem[Solar-Lezama et~al\mbox{.}(2008)]%
        {psketch}
\bibfield{author}{\bibinfo{person}{Armando Solar-Lezama}, \bibinfo{person}{Christopher~Grant Jones}, {and} \bibinfo{person}{Rastislav Bodik}.} \bibinfo{year}{2008}\natexlab{}.
\newblock \showarticletitle{Sketching Concurrent Data Structures}. In \bibinfo{booktitle}{\emph{{PLDI}}}.
\newblock
\href{https://doi.org/10.1145/1375581.1375599}{doi:\nolinkurl{10.1145/1375581.1375599}}


\bibitem[Solar-Lezama et~al\mbox{.}(2006)]%
        {Solar-LezamaETALCombSketchFinite2006}
\bibfield{author}{\bibinfo{person}{Armando Solar-Lezama}, \bibinfo{person}{Liviu Tancau}, \bibinfo{person}{Rastislav Bodik}, \bibinfo{person}{Sanjit Seshia}, {and} \bibinfo{person}{Vijay Saraswat}.} \bibinfo{year}{2006}\natexlab{}.
\newblock \showarticletitle{Combinatorial Sketching for Finite Programs}. In \bibinfo{booktitle}{\emph{ASPLOS}}.
\newblock
\href{https://doi.org/10.1145/1168857.1168907}{doi:\nolinkurl{10.1145/1168857.1168907}}


\bibitem[Sun et~al\mbox{.}(2023)]%
        {mate}
\bibfield{author}{\bibinfo{person}{Ruoxi Sun}, \bibinfo{person}{Minhui Xue}, \bibinfo{person}{Gareth Tyson}, \bibinfo{person}{Tian Dong}, \bibinfo{person}{Shaofeng Li}, \bibinfo{person}{Shuo Wang}, \bibinfo{person}{Haojin Zhu}, \bibinfo{person}{Seyit Camtepe}, {and} \bibinfo{person}{Surya Nepal}.} \bibinfo{year}{2023}\natexlab{}.
\newblock \showarticletitle{Mate! Are You Really Aware? An Explainability-Guided Testing Framework for Robustness of Malware Detectors} \emph{(\bibinfo{series}{ESEC/FSE 2023})}.
\newblock
\href{https://doi.org/10.1145/3611643.3616309}{doi:\nolinkurl{10.1145/3611643.3616309}}


\bibitem[Thimmaiah et~al\mbox{.}(2025)]%
        {ThimmaiahETAL25eStore}
\bibfield{author}{\bibinfo{person}{Aditya Thimmaiah}, \bibinfo{person}{Zijian Yi}, \bibinfo{person}{Joseph Kenis}, \bibinfo{person}{Christopher~J. Rossbach}, {and} \bibinfo{person}{Milos Gligoric}.} \bibinfo{year}{2025}\natexlab{}.
\newblock \showarticletitle{In-memory Object Graph Stores}. In \bibinfo{booktitle}{\emph{European Conference on Object-Oriented Programming}}. \bibinfo{pages}{30:1--30:30}.
\newblock
\href{https://doi.org/10.4230/LIPIcs.ECOOP.2025.30}{doi:\nolinkurl{10.4230/LIPIcs.ECOOP.2025.30}}


\bibitem[Wadhwa et~al\mbox{.}(2024)]%
        {WadhwaETALFSE2024}
\bibfield{author}{\bibinfo{person}{Nalin Wadhwa}, \bibinfo{person}{Jui Pradhan}, \bibinfo{person}{Atharv Sonwane}, \bibinfo{person}{Surya~Prakash Sahu}, \bibinfo{person}{Nagarajan Natarajan}, \bibinfo{person}{Aditya Kanade}, \bibinfo{person}{Suresh Parthasarathy}, {and} \bibinfo{person}{Sriram Rajamani}.} \bibinfo{year}{2024}\natexlab{}.
\newblock \showarticletitle{{CORE}: Resolving Code Quality Issues using {LLMs}}.
\newblock  \bibinfo{number}{FSE} (\bibinfo{year}{2024}).
\newblock
\href{https://doi.org/10.1145/3643762}{doi:\nolinkurl{10.1145/3643762}}


\bibitem[Wang et~al\mbox{.}(2023a)]%
        {wang2023mlirsmith}
\bibfield{author}{\bibinfo{person}{Haoyu Wang}, \bibinfo{person}{Junjie Chen}, \bibinfo{person}{Chuyue Xie}, \bibinfo{person}{Shuang Liu}, \bibinfo{person}{Zan Wang}, \bibinfo{person}{Qingchao Shen}, {and} \bibinfo{person}{Yingquan Zhao}.} \bibinfo{year}{2023}\natexlab{a}.
\newblock \showarticletitle{Mlirsmith: Random program generation for fuzzing mlir compiler infrastructure}. In \bibinfo{booktitle}{\emph{ASE}}.
\newblock
\href{https://doi.org/10.1109/ASE56229.2023.00120}{doi:\nolinkurl{10.1109/ASE56229.2023.00120}}


\bibitem[Wang et~al\mbox{.}(2019)]%
        {wang2019superion}
\bibfield{author}{\bibinfo{person}{Junjie Wang}, \bibinfo{person}{Bihuan Chen}, \bibinfo{person}{Lei Wei}, {and} \bibinfo{person}{Yang Liu}.} \bibinfo{year}{2019}\natexlab{}.
\newblock \showarticletitle{Superion: Grammar-aware greybox fuzzing}. In \bibinfo{booktitle}{\emph{ICSE}}.
\newblock
\href{https://doi.org/10.1109/icse.2019.00081}{doi:\nolinkurl{10.1109/icse.2019.00081}}


\bibitem[Wang et~al\mbox{.}(2023b)]%
        {fuzzjit}
\bibfield{author}{\bibinfo{person}{Junjie Wang}, \bibinfo{person}{Zhiyi Zhang}, \bibinfo{person}{Shuang Liu}, \bibinfo{person}{Xiaoning Du}, {and} \bibinfo{person}{Junjie Chen}.} \bibinfo{year}{2023}\natexlab{b}.
\newblock \showarticletitle{FuzzJIT: Oracle-enhanced fuzzing for JavaScript engine JIT compiler}. In \bibinfo{booktitle}{\emph{USENIX Security}} \emph{(\bibinfo{series}{SEC '23})}.
\newblock
\urldef\tempurl%
\url{https://www.usenix.org/conference/usenixsecurity23/presentation/wang-junjie}
\showURL{%
\tempurl}


\bibitem[Wang et~al\mbox{.}(2018)]%
        {WangETALABZ2018ASketch}
\bibfield{author}{\bibinfo{person}{Kaiyuan Wang}, \bibinfo{person}{Allison Sullivan}, \bibinfo{person}{Darko Marinov}, {and} \bibinfo{person}{Sarfraz Khurshid}.} \bibinfo{year}{2018}\natexlab{}.
\newblock \showarticletitle{Solver-based Sketching of {Alloy} Models using Test Valuations}. In \bibinfo{booktitle}{\emph{ABZ}}.
\newblock
\href{https://doi.org/10.1007/978-3-319-91271-4_9}{doi:\nolinkurl{10.1007/978-3-319-91271-4_9}}


\bibitem[Wang et~al\mbox{.}(2024)]%
        {WangETAL2024}
\bibfield{author}{\bibinfo{person}{Wei Wang}, \bibinfo{person}{Huilong Ning}, \bibinfo{person}{Gaowei Zhang}, \bibinfo{person}{Libo Liu}, {and} \bibinfo{person}{Yi Wang}.} \bibinfo{year}{2024}\natexlab{}.
\newblock \showarticletitle{Rocks Coding, Not Development: A Human-Centric, Experimental Evaluation of {LLM}-Supported {SE} Tasks}.
\newblock  \bibinfo{number}{FSE} (\bibinfo{year}{2024}).
\newblock
\href{https://doi.org/10.1145/3643758}{doi:\nolinkurl{10.1145/3643758}}


\bibitem[Xia et~al\mbox{.}(2025)]%
        {XiaETALFSE2025}
\bibfield{author}{\bibinfo{person}{Chunqiu~Steven Xia}, \bibinfo{person}{Yinlin Deng}, \bibinfo{person}{Soren Dunn}, {and} \bibinfo{person}{Lingming Zhang}.} \bibinfo{year}{2025}\natexlab{}.
\newblock \showarticletitle{Agentless: Demystifying {LLM}-based Software Engineering Agents}. In \bibinfo{booktitle}{\emph{{ESEC/FSE}}}. \bibinfo{numpages}{24}~pages.
\newblock
\href{https://doi.org/10.1145/3715754}{doi:\nolinkurl{10.1145/3715754}}


\bibitem[Xia et~al\mbox{.}(2024)]%
        {xia2024fuzz4all}
\bibfield{author}{\bibinfo{person}{Chunqiu~Steven Xia}, \bibinfo{person}{Matteo Paltenghi}, \bibinfo{person}{Jia Le~Tian}, \bibinfo{person}{Michael Pradel}, {and} \bibinfo{person}{Lingming Zhang}.} \bibinfo{year}{2024}\natexlab{}.
\newblock \showarticletitle{{Fuzz4All}: Universal Fuzzing with Large Language Models}. In \bibinfo{booktitle}{\emph{International Conference on Software Engineering}}. Article \bibinfo{articleno}{126}, \bibinfo{numpages}{13}~pages.
\newblock


\bibitem[Xie et~al\mbox{.}(2023)]%
        {10.1145/3597926.3598061}
\bibfield{author}{\bibinfo{person}{Zifan Xie}, \bibinfo{person}{Ming Wen}, \bibinfo{person}{Haoxiang Jia}, \bibinfo{person}{Xiaochen Guo}, \bibinfo{person}{Xiaotong Huang}, \bibinfo{person}{Deqing Zou}, {and} \bibinfo{person}{Hai Jin}.} \bibinfo{year}{2023}\natexlab{}.
\newblock \showarticletitle{Precise and Efficient Patch Presence Test for Android Applications against Code Obfuscation} \emph{(\bibinfo{series}{ISSTA 2023})}.
\newblock
\href{https://doi.org/10.1145/3597926.3598061}{doi:\nolinkurl{10.1145/3597926.3598061}}


\bibitem[Xu et~al\mbox{.}(2024)]%
        {XuETALICSE2024}
\bibfield{author}{\bibinfo{person}{Junjielong Xu}, \bibinfo{person}{Ziang Cui}, \bibinfo{person}{Yuan Zhao}, \bibinfo{person}{Xu Zhang}, \bibinfo{person}{Shilin He}, \bibinfo{person}{Pinjia He}, \bibinfo{person}{Liqun Li}, \bibinfo{person}{Yu Kang}, \bibinfo{person}{Qingwei Lin}, \bibinfo{person}{Yingnong Dang}, \bibinfo{person}{Saravan Rajmohan}, {and} \bibinfo{person}{Dongmei Zhang}.} \bibinfo{year}{2024}\natexlab{}.
\newblock \showarticletitle{UniLog: Automatic Logging via {LLM} and In-Context Learning}. In \bibinfo{booktitle}{\emph{ICSE}}.
\newblock
\href{https://doi.org/10.1145/3597503.3623326}{doi:\nolinkurl{10.1145/3597503.3623326}}


\bibitem[Yang et~al\mbox{.}(2011)]%
        {csmith}
\bibfield{author}{\bibinfo{person}{Xuejun Yang}, \bibinfo{person}{Yang Chen}, \bibinfo{person}{Eric Eide}, {and} \bibinfo{person}{John Regehr}.} \bibinfo{year}{2011}\natexlab{}.
\newblock \showarticletitle{Finding and understanding bugs in C compilers}. In \bibinfo{booktitle}{\emph{PLDI}}.
\newblock
\href{https://doi.org/10.1145/1993316.1993532}{doi:\nolinkurl{10.1145/1993316.1993532}}


\bibitem[Yang et~al\mbox{.}(2018)]%
        {YangETAL2017}
\bibfield{author}{\bibinfo{person}{Zijiang Yang}, \bibinfo{person}{Jinru Hua}, \bibinfo{person}{Kaiyuan Wang}, {and} \bibinfo{person}{Sarfraz Khurshid}.} \bibinfo{year}{2018}\natexlab{}.
\newblock \showarticletitle{EdSynth: Synthesizing {API} Sequences with Conditionals and Loops}. In \bibinfo{booktitle}{\emph{{ICST}}}.
\newblock
\href{https://doi.org/10.1109/ICST.2018.00025}{doi:\nolinkurl{10.1109/ICST.2018.00025}}


\bibitem[Yu et~al\mbox{.}(2025)]%
        {yu2025ratte}
\bibfield{author}{\bibinfo{person}{Pingshi Yu}, \bibinfo{person}{Nicolas Wu}, {and} \bibinfo{person}{Alastair~F Donaldson}.} \bibinfo{year}{2025}\natexlab{}.
\newblock \showarticletitle{Ratte: Fuzzing for Miscompilations in Multi-Level Compilers Using Composable Semantics}. In \bibinfo{booktitle}{\emph{ASPLOS}}.
\newblock
\href{https://doi.org/10.1145/3676641.3716270}{doi:\nolinkurl{10.1145/3676641.3716270}}


\bibitem[Yuan et~al\mbox{.}(2024)]%
        {10795040}
\bibfield{author}{\bibinfo{person}{Hengchen Yuan}, \bibinfo{person}{Jiefang Lin}, \bibinfo{person}{Wing Lam}, {and} \bibinfo{person}{August Shi}.} \bibinfo{year}{2024}\natexlab{}.
\newblock \showarticletitle{{ Test Scheduling Across Heterogeneous Machines While Balancing Running Time, Price, and Flakiness }}. In \bibinfo{booktitle}{\emph{ICSME}}.
\newblock
\href{https://doi.org/10.1109/ICSME58944.2024.00048}{doi:\nolinkurl{10.1109/ICSME58944.2024.00048}}


\bibitem[Yuan et~al\mbox{.}(2023)]%
        {yuan2023well}
\bibfield{author}{\bibinfo{person}{Zheng Yuan}, \bibinfo{person}{Hongyi Yuan}, \bibinfo{person}{Chuanqi Tan}, \bibinfo{person}{Wei Wang}, {and} \bibinfo{person}{Songfang Huang}.} \bibinfo{year}{2023}\natexlab{}.
\newblock \showarticletitle{How well do large language models perform in arithmetic tasks?}
\newblock \bibinfo{journal}{\emph{arXiv preprint arXiv:2304.02015}} (\bibinfo{year}{2023}).
\newblock
\href{https://doi.org/10.48550/arXiv.2304.02015}{doi:\nolinkurl{10.48550/arXiv.2304.02015}}


\bibitem[Zang et~al\mbox{.}(2024a)]%
        {zang2024jog}
\bibfield{author}{\bibinfo{person}{Zhiqiang Zang}, \bibinfo{person}{Aditya Thimmaiah}, {and} \bibinfo{person}{Milos Gligoric}.} \bibinfo{year}{2024}\natexlab{a}.
\newblock \showarticletitle{JOG: Java JIT peephole optimizations and tests from patterns}. In \bibinfo{booktitle}{\emph{ICSE}}.
\newblock
\href{https://doi.org/10.1145/3639478.3640040}{doi:\nolinkurl{10.1145/3639478.3640040}}


\bibitem[Zang et~al\mbox{.}(2024b)]%
        {lejit}
\bibfield{author}{\bibinfo{person}{Zhiqiang Zang}, \bibinfo{person}{Fu-Yao Yu}, \bibinfo{person}{Aditya Thimmaiah}, \bibinfo{person}{August Shi}, {and} \bibinfo{person}{Milos Gligoric}.} \bibinfo{year}{2024}\natexlab{b}.
\newblock \showarticletitle{Java JIT Testing with Template Extraction}.
\newblock  \bibinfo{number}{FSE} (\bibinfo{year}{2024}).
\newblock
\href{https://doi.org/10.1145/3643777}{doi:\nolinkurl{10.1145/3643777}}


\bibitem[Zang et~al\mbox{.}(2023)]%
        {jattack}
\bibfield{author}{\bibinfo{person}{Zhiqiang Zang}, \bibinfo{person}{Fu-Yao Yu}, \bibinfo{person}{Nathan Wiatrek}, \bibinfo{person}{Milos Gligoric}, {and} \bibinfo{person}{August Shi}.} \bibinfo{year}{2023}\natexlab{}.
\newblock \showarticletitle{JAttack: Java JIT Testing Using Template Programs} \emph{(\bibinfo{series}{ICSE '23})}.
\newblock
\href{https://doi.org/10.1109/ICSE-Companion58688.2023.00014}{doi:\nolinkurl{10.1109/ICSE-Companion58688.2023.00014}}


\bibitem[Zhang et~al\mbox{.}(2019)]%
        {10.1145/3293882.3330563}
\bibfield{author}{\bibinfo{person}{Jiexin Zhang}, \bibinfo{person}{Alastair~R. Beresford}, {and} \bibinfo{person}{Stephan~A. Kollmann}.} \bibinfo{year}{2019}\natexlab{}.
\newblock \showarticletitle{LibID: reliable identification of obfuscated third-party Android libraries} \emph{(\bibinfo{series}{ISSTA 2019})}.
\newblock
\href{https://doi.org/10.1145/3293882.3330563}{doi:\nolinkurl{10.1145/3293882.3330563}}


\bibitem[Zhang et~al\mbox{.}(2017)]%
        {Skeletal2017}
\bibfield{author}{\bibinfo{person}{Qirun Zhang}, \bibinfo{person}{Chengnian Sun}, {and} \bibinfo{person}{Zhendong Su}.} \bibinfo{year}{2017}\natexlab{}.
\newblock \showarticletitle{Skeletal program enumeration for rigorous compiler testing}. In \bibinfo{booktitle}{\emph{PLDI}}.
\newblock
\href{https://doi.org/10.1145/3140587.3062379}{doi:\nolinkurl{10.1145/3140587.3062379}}


\bibitem[Zhang et~al\mbox{.}(2021)]%
        {zhang2021android}
\bibfield{author}{\bibinfo{person}{Xiaolu Zhang}, \bibinfo{person}{Frank Breitinger}, \bibinfo{person}{Engelbert Luechinger}, {and} \bibinfo{person}{Stephen O'Shaughnessy}.} \bibinfo{year}{2021}\natexlab{}.
\newblock \showarticletitle{Android application forensics: A survey of obfuscation, obfuscation detection and deobfuscation techniques and their impact on investigations}.
\newblock \bibinfo{journal}{\emph{Forensic Science International: Digital Investigation}} (\bibinfo{year}{2021}).
\newblock
\href{https://doi.org/10.1016/j.fsidi.2021.301285}{doi:\nolinkurl{10.1016/j.fsidi.2021.301285}}


\bibitem[Zhao et~al\mbox{.}(2024)]%
        {zhao2024docmath}
\bibfield{author}{\bibinfo{person}{Yilun Zhao}, \bibinfo{person}{Yitao Long}, \bibinfo{person}{Hongjun Liu}, \bibinfo{person}{Ryo Kamoi}, \bibinfo{person}{Linyong Nan}, \bibinfo{person}{Lyuhao Chen}, \bibinfo{person}{Yixin Liu}, \bibinfo{person}{Xiangru Tang}, \bibinfo{person}{Rui Zhang}, {and} \bibinfo{person}{Arman Cohan}.} \bibinfo{year}{2024}\natexlab{}.
\newblock \showarticletitle{Docmath-eval: Evaluating math reasoning capabilities of llms in understanding financial documents}. In \bibinfo{booktitle}{\emph{ACL}}.
\newblock
\href{https://doi.org/10.18653/v1/2024.acl-long.852}{doi:\nolinkurl{10.18653/v1/2024.acl-long.852}}


\bibitem[Zhong et~al\mbox{.}(2025a)]%
        {zhong2025approach}
\bibfield{author}{\bibinfo{person}{Hua Zhong}, \bibinfo{person}{Shan Jiang}, {and} \bibinfo{person}{Sarfraz Khurshid}.} \bibinfo{year}{2025}\natexlab{a}.
\newblock \showarticletitle{An approach for API synthesis using large language models}.
\newblock \bibinfo{journal}{\emph{arXiv preprint arXiv:2502.15246}} (\bibinfo{year}{2025}).
\newblock
\href{https://doi.org/10.48550/arXiv.2502.15246}{doi:\nolinkurl{10.48550/arXiv.2502.15246}}


\bibitem[Zhong et~al\mbox{.}(2025b)]%
        {zhong2025april}
\bibfield{author}{\bibinfo{person}{Hua Zhong}, \bibinfo{person}{Shan Jiang}, {and} \bibinfo{person}{Sarfraz Khurshid}.} \bibinfo{year}{2025}\natexlab{b}.
\newblock \showarticletitle{APRIL: API Synthesis with Automatic Prompt Optimization and Reinforcement Learning}.
\newblock \bibinfo{journal}{\emph{arXiv preprint arXiv:2509.25196}} (\bibinfo{year}{2025}).
\newblock
\href{https://doi.org/10.48550/arXiv.2509.25196}{doi:\nolinkurl{10.48550/arXiv.2509.25196}}


\end{thebibliography}

\end{document}